\documentclass[12pt]{article}
\usepackage{latexsym}
\usepackage{amsmath,,calrsfs}
\usepackage{amsfonts}
\usepackage{amssymb}
\usepackage{amscd}
\usepackage{bbm}
\usepackage{fancybox}
\usepackage{cite}
\usepackage{amsmath,amsfonts,amsbsy}
\usepackage{pstricks,pst-node}
\usepackage[small,bf,hang]{caption2}
\usepackage{graphicx}
\usepackage{epsfig}
\usepackage{psfrag}
\usepackage{comment}

\usepackage{float}

\psset{unit=1.3cm,linewidth=.5pt,radius=.2}  

\usepackage{multirow}                     
\usepackage{float}                          
\usepackage{lscape}                         
\usepackage{bm}

\newcommand{\beq}{\begin{equation}}
\newcommand{\eeq}{\end{equation}}
\newcommand{\bi}{\begin{itemize}}
\newcommand{\ei}{\end{itemize}}
\newcommand{\bt}{\begin{tabular}}
\newcommand{\et}{\end{tabular}}
\newcommand{\bc}{\begin{center}}
\newcommand{\ec}{\end{center}}

\newcommand{\be}{\begin{equation}}
\newcommand{\ee}{\end{equation}}
\newcommand{\bea}{\begin{eqnarray}}
\newcommand{\eea}{\end{eqnarray}}
\newcommand{\ba}{\begin{array}}
\newcommand{\ea}{\end{array}}
\newcommand{\p}[1]{(\ref{#1})}
\newcommand{\lb}[1]{\label{#1}}

\def\bbox{{\,\lower0.9pt\vbox{\hrule \hbox{\vrule height 0.2 cm
\hskip 0.2 cm \vrule height 0.2 cm}\hrule}\,}}
\newcommand{\dsl}{\pa \kern-0.5em /}

\newcommand{\nn}{\nonumber \\}

\makeatletter \@addtoreset{equation}{section} \makeatother

\def\slashchar#1{\setbox0=\hbox{$#1$}           
   \dimen0=\wd0                                 
   \setbox1=\hbox{/} \dimen1=\wd1               
   \ifdim\dimen0>\dimen1                        
      \rlap{\hbox to \dimen0{\hfil/\hfil}}      
      #1                                        
   \else                                        
      \rlap{\hbox to \dimen1{\hfil$#1$\hfil}}   
      /                                         
   \fi}

\begin{document}

\begin{titlepage}

\renewcommand{\thefootnote}{\star}

\begin{center}

\hfill  {}


{\Large \bf  Global Symmetries of Quaternion-K$\bf \ddot{a}$hler
\vspace{0.3cm}

${\cal N}=4$ Supersymmetric Mechanics}

\vspace{0.3cm}

\vspace{1.3cm}
\renewcommand{\thefootnote}{$\star$}

{\large\bf Evgeny~Ivanov}${}^{a, b}$,
{\quad \large\bf Luca~Mezincescu}${}^c$
 \vspace{0.5cm}

{${}^a$\it Bogoliubov Laboratory of Theoretical Physics, JINR,}\\
{\it 141980 Dubna, Moscow region, Russia} \\
{${}^b$\it Moscow Institute of Physics and Technology,}\\
{\it141700 Dolgoprudny,  Moscow region, Russia}\\
\vspace{0.1cm}

{\tt eivanov@theor.jinr.ru}\\
\vspace{0.4cm}

{${}^c$\it Department of Physics, University of Miami,}\\
{\it P.O. Box 248046, Coral Gables,
FL 33124, USA}\\
\vspace{0.1cm}

{\tt  mezincescu@physics.miami.edu}\\

\end{center}
\vspace{0.2cm} \vskip 0.6truecm \nopagebreak

\begin{abstract}
\noindent We analyze the global symmetries of ${\cal N}=4$ supersymmetric mechanics
involving $4n$-dimensional Quaternion-K\"ahler (QK) $1D$ sigma models on projective spaces
$\mathbb{H}{\rm H}^n$ and $\mathbb{H}{\rm P}^n$ as the bosonic core.
All Noether charges associated with global worldline symmetries are shown to vanish as a result of equations of motion, which implies that we deal
with a severely constrained hamiltonian system. The complete hamiltonian analysis
of the bosonic sector is performed.
\end{abstract}
\vskip 1cm

\vskip 0.5cm

\vspace{1cm}

\phantom{Keywords: }

\newpage

\end{titlepage}

\setcounter{footnote}{0}

\newpage
\setcounter{page}{1}

\section{Introduction}
After appearance of the first versions of supersymmetric quantum
mechanics (SQM) based on ${\cal N}=2$ one-dimensional supersymmetry
\cite{Nic}, \cite{Witten}, this class of theories was intensively
and extensively studied in numerous articles and reviews (see, e.g.,
\cite{cole} - \cite{FIL}). The SQM models reveal interesting quantum
and geometric properties, some of which cannot be reproduced in the
framework of the standard dimensional reduction from the
higher-dimensional supersymmetric field theories. Of special
interest are SQM models with extended ${\cal N}=4$ and ${\cal N}=8$
worldline supersymmetries. Some of these $1D$ sigma models admit,
e.g., hyper-K\"ahler manifolds as their bosonic target spaces and
are capable to provide a good laboratory for analyzing various
properties of the higher-dimensional theories associated with such
kind of targets.

Until recently, the majority of SQM models (including those  with ${\cal N} \geq 4$) were constructed under an assumption that the worldline
(or super worldline in the case of superfield models)
are ``flat'', i.e. $1D$ supersymmetry is rigid. In the paper \cite{Ivanov:2017ajf}, using ${\cal N}=4, 1D$ harmonic superspace approach \cite{IvLe}, we constructed
a new type of ${\cal N}=4$ supersymmetric mechanics involving $4n$-dimensional Quaternion-K\"ahler (QK) $1D$ sigma models as the bosonic core. The basic distinguishing
features of the new SQM models constructed are local worldline ${\cal N}=4$ supersymmetry and the presence of the appropriate ${\cal N}=4, 1D$ supergravity
multiplet for ensuring this local invariance.

When restricted to the $\mathbb{H}{\rm H}^n$ or $\mathbb{H}{\rm P}^n $ target manifolds, the bosonic sector of these models
is identical to the dimensionally reduced homogeneous QK sigma models of refs. \cite{QK}, \cite{Gal}, \cite {Ivanov:1999vg}. In ref. \cite{Ivanov:2017ajf}
the component action was also obtained for the fermionic sector, therefore the whole component action in an arbitrary gauge
is available. With so much detailed information accessible it is worth trying some further insights into these models.
These models have rather large gauge symmetries, so that the quantization in a covariant manner seems challenging.
In this paper we will restrict ourselves to the study of the global symmetries which these models possess,
restricting the gauge transformations to constant parameters.

We will start by recalling the pivotal points of the harmonic superfield formulation of general QK ${\cal N}=4$ supersymmetric mechanics
and presenting a short review of the QK component $\mathbb{H}{\rm H}^n$ and $\mathbb{H}{\rm P}^n $ models followed by a review of their transformation properties. Then we will concentrate
on the global properties of these models, restricting the gauge transformations to constant parameters. After
a somehow involved algebra we will show  that the corresponding Noether charges are vanishing as a consequence of the equations of motion,
and therefore give rise to the gauge constraints, in a similar fashion, e.g., to the appearance of Virasoro constraints
in the string theory. The hamiltonian analysis of the bosonic sectors of these homogeneous models is performed
as a prerequisite to their quantization.

\section{QK supersymmetric mechanics in superspace}
Here we recall the basic points of the construction in ref.\cite{Ivanov:2017ajf}.

\subsection{${\cal N} = 4\,, \,1D$ harmonic superspace setup}

One-dimensional ${\cal N}=4$ supersymmetry admits a natural realization in
$1D$ harmonic superspace \cite{IvLe}.
In the analytic basis, this superspace is represented by the coordinate set
\bea
z :=
(t_A, \theta^+, \bar\theta^+, \theta^-, \bar\theta^-, w^\pm_i)\,,
\quad w^+_iw^-_k - w^+_k w^-_i = \varepsilon_{ik}\,,
\lb{1DHSS}
\eea
where $w^\pm_i$  are harmonics parametrizing the automorphism $SU(2)$ group. The analytic harmonic superspace is a subset of
\p{1DHSS}
\bea
\zeta := (t_A, \theta^+, \bar\theta^+, w^\pm_i)\,.
\lb{1DASS}
\eea
In what follows, we will omit the subscript $A$ of $t_A$. Both \p{1DHSS} and \p{1DASS} are closed under the
appropriate realization of ${\cal N}=4$ supersymmetry \footnote{The ${\cal N}=4$ transformations of the superspace coordinates, as well as the relations between
the analytic and central bases of harmonic superspace can be found in \cite{IvLe}
and in a more recent paper \cite{DI1}.}.  These coordinate sets are also closed and real with
respect to the generalized $\,\widetilde{\,}\,$ - conjugation,
\bea
\widetilde{t} = t\,, \quad \widetilde{\theta^{\pm}} =
\bar\theta^{\pm}\,, \; \widetilde{\bar\theta^{\pm}} = -
\theta^{\pm}\,, \quad \widetilde{w^\pm_i} = w^{\pm\,i} =
\varepsilon^{ik}{w^\pm_k}\,, \; \widetilde{w^{\pm\,i}} = -
w^{\pm}_{i}\,.\lb{Tilde}
\eea

An important ingredient of the harmonic superspace formalism is the harmonic derivatives
\bea
&& D^{++} = \partial^{++} + 2i\theta^+\bar\theta^+ \partial_t + \theta^+\partial_{\theta^-} + \bar\theta^+\partial_{\bar\theta^-}\,, \lb{D++} \\
&& D^{--} =
\partial^{--} + 2i\theta^-\bar\theta^- \partial_t + \theta^-\partial_{\theta^+}
+ \bar\theta^-\partial_{\bar\theta^+}\,, \lb{D--} \\
&& [D^{++}, D^{--}] = D^0 = \partial^0 + \theta^+\partial_{\theta^+}
+ \bar\theta^+\partial_{\bar\theta^+} - \theta^-\partial_{\theta^-}
- \bar\theta^-\partial_{\bar\theta^-}\,, \lb{DDcom} \eea where \be
\partial^{\pm\pm} = w^\pm_i \frac{\partial}{\partial w^\mp_i}\,, \quad \partial^0 =
w^+_i \frac{\partial}{\partial w^+_i} - w^-_i
\frac{\partial}{\partial w^-_i}\,.
\ee
The covariant derivative $D^{++}$ is distinguished in that it preserves the analyticity: the result of its action on the
analytic superfield $\Phi(\zeta)$, i.e. $D^{++}\Phi(\zeta)$, is
again an analytic $1D$ superfield. The operator
$D^0$ counts the external harmonic $U(1)$ charges of the
harmonic $1D$ superfields. All superfields
are assumed to possess a definite harmonic $U(1)$ charge and so are
eigenfunctions of $D^0$.

The QK supersymmetric mechanics is formulated in terms of two analytic superfields $q^{+
a}(\zeta)$ and $2n$ analytic superfields $\hat{Q}^{+ r}(\zeta)$,
where $a=1,2$ and $r = 1, \ldots 2n$ are, respectively, the indices
of the fundamental representations of some extra groups $Sp(1) \sim
SU(2)$ and $Sp(n)$ commuting with supersymmetry. They
are subjected to the tilde-reality conditions
\bea
\widetilde{q^+_a}
= \varepsilon^{ab}q^+_b, \quad \widetilde{\hat{Q}^+_r} =
\Omega^{rs}\,\hat{Q}^+_s\,, \lb{Reality}
\eea
where $\Omega^{rs},
\Omega^{rp}\Omega_{ps} = \delta^r_s\,,$ are skew-symmetric constant
$Sp(n)$ invariant ``metrics''. The superfields $q^+_a$ and $\hat{Q}^+_s$ are $1D$
analogs of the compensating hypermultiplet and ``matter''
hypermultiplets \cite{SGN2,HSS,GIOO,Ivanov:1999vg}.

In the case of simplest
$\mathbb{H}{\rm H}^n$ and $\mathbb{H}{\rm P}^n$ sigma models which
will be the main subject of the present paper,
these superfields are subject to the linear harmonic constraints
\bea
D^{++}q^+_a(\zeta) = 0\,, \qquad  D^{++} \hat{Q}^+_r(\zeta) = 0\,. \lb{HSconstrLin}
\eea
The general QK sigma model amounts to some nonlinear version of these constraints
\bea
&& D^{++} q^{+ a} - \gamma \frac12 \frac{\partial}{\partial q^{+}_a}\Big[\hat{\kappa}^2 (w^-\cdot q^+)^2{\cal L}^{+4} \Big] = 0\,, \quad \gamma = \pm 1\,, \lb{qgen} \\
&& D^{++} \hat{Q}^{+ r} + \frac12 \frac{\partial}{\partial \hat{Q}^{+}_r}\Big[\hat{\kappa}^2 (w^-\cdot q^+)^2{\cal L}^{+4} \Big]
= 0\,,\lb{Qgen} \\
&& {\cal L}^{+4} \equiv {\cal L}^{+4} \Big(\frac{\hat{Q}^{ + r}}{\hat{\kappa} (w^-\cdot q^+)}, \frac{q^{ + a}}{(w^-\cdot q^+)}, w^-_i\Big),
\quad (w^-\cdot q^+) := w^-_{a}q^{+ a}\,. \lb{L+4}
\eea
The object ${\cal L}^{+4}$ is just the QK potential introduced in \cite{SGN2} and derived in \cite{GIOO} within the pure geometric setting as the fundamental object of QK geometry;
the parameter $\hat\kappa$ is the contraction parameter to the general HK
case: when $\hat\kappa$ goes to zero, $\hat{\kappa} (w^-\cdot q^+) \rightarrow 1\,, \; \frac{q^{ + a}}{(w^-\cdot q^+)} \rightarrow w^{+a}$
and,
respectively, ${\cal L}^{+4} \rightarrow {\cal L}^{+4}(Q, w^+, w^-)$. In this limit \p{Qgen} becomes the nonlinear constraint describing
the most general HK ${\cal N}=4, 1D$ sigma model \cite{DI1}. The superfield $q^+_a$ fully decouples in this limit. It is important that the QK potential
${\cal L}^{+4}$ originally does not involve any explicit $w^+$ harmonics. The parameter $\gamma = \pm 1$ in \p{qgen} discriminates the cases with non-compact
$\mathbb{H}{\rm H} = Sp(n,1)/[Sp(1) \times Sp(n)]$ ($\gamma = 1$) and  compact $\mathbb{H}{\rm P} = Sp(n+1)/[Sp(1) \times Sp(n)]$ ($\gamma = -1$) homogeneous
projective manifolds as the ``flattest'' $n$-dimensional QK ones (see next Subsection).

While solving the generic nonlinear constraints \p{qgen}, \p{Qgen} is a rather involved problem, it is much easier for their linear version \p{HSconstrLin}. The
explicit solution of \p{HSconstrLin} is as follows
\bea && q^{+\,
a}(\zeta) = f^{ia}(t)w^+_i + \theta^+\chi^a(t) -
\bar\theta^+\bar\chi^a(t) - 2i\theta^+\bar\theta^+
\dot{f}^{ia}(t)w^-_i\,,
\lb{qcomp} \\
&& \hat{Q}^{+\, r}(\zeta) = \hat{F}^{ir}(t)w^+_i + \theta^+\chi^r(t)
- \bar\theta^+\bar\chi^r(t) - 2i\theta^+\bar\theta^+
\dot{\hat{F}}^{ir}(t)w^-_i\,. \lb{Qcomp}
\eea
The superfield reality conditions \p{Reality} imply the following reality properties for
the component fields:
\bea
\widetilde{f^+_a} = f^{+ a}
\,\Leftrightarrow \, \overline{(f_{ia})} = f^{ia}\,, \;
\overline{(f^{ia})} = f_{ia}\,; \qquad \overline{(\chi_a)} =
\bar{\chi}^a\,, \;\overline{(\chi^a)} = -\bar{\chi}_a \label{qQreal}
\eea
(and similar ones for $\hat{F}^{ir}, \chi^r$). It is assumed that the
indices $a$ and $r$ are raised and lowered in the standard way by
the skew-symmetric tensors $\varepsilon_{ab}, \varepsilon^{ab}$ and
$\Omega_{rs}, \Omega^{rs}$. We observe that $q^{+ a}$ carries 4 real
bosonic degrees of freedom and $\hat{Q}^{+\, r}$ - $4n$ such
degrees, total of $4(n+1)$ bosonic degrees.

One more relevant superfield (just specific for QK models) is the scalar real superfield
$E(z)$ which collects the objects of ${\cal N}=4, 1D$
``supergravity''. It lives on the whole harmonic superspace
\p{1DHSS} and is subjected to the purely harmonic constraint
\be
D^{++}E = 0\,, \quad \widetilde{E} = E\,, \lb{Hconstr}
\ee
which means that $E$ in the central basis does not depend   on
harmonics at all, and so is the standard harmonic-independent ${\cal
N}=4\,, 1D$ superfield. In the analytic basis, the component
structure of $E$ is as follows
\bea
E(z) &=& h + \theta^+\theta^- M
- \bar\theta^+\bar\theta^- \bar M + \theta^+\bar\theta^- (\mu -
i\dot{h}) + \bar\theta^+\theta^- (\mu + i\dot{h}) \nn && + \,4i
(\theta^+\bar\theta^+ w^-_iw^-_k - \theta^+\bar\theta^- w^-_iw^+_k -
\theta^-\bar\theta^+ w^-_iw^+_k + \theta^-\bar\theta^- w^+_iw^+_k)
L^{(ik)} \nn && +\,4\theta^+\bar\theta^+\theta^-\bar\theta^- [D +
2\dot{L}^{(ik)}w^+_iw^-_k]
\lb{BosH} \\
&& +\, (\theta^-w^+_i - \theta^+w^-_i)\phi^i - (\bar\theta^-w^+_i -
\bar\theta^+w^-_i)\bar\phi^i + 4i \theta^-\bar\theta^-
(\theta^+w^+_i \sigma^i - \bar\theta^+w^+_i \bar\sigma^i) \nn && +
\,2i\theta^+\bar\theta^+ [\theta^-w^-_i(2\sigma^i - \dot{\phi}^i)
-\bar\theta^-w^-_i(2\bar\sigma^i - \dot{\bar\phi}^i)]\,.\lb{FermH}
\eea
It includes eight bosonic fields $h(t), M(t), \bar M(t),
\mu(t), D(t), L^{(ik)}(t)$ and eight fermionic fields $\phi^i(t),
\bar\phi^i(t), \sigma^i(t), \bar\sigma^i(t)$. The conjugation rules
for all these fields read
\bea
&& \overline{(h)} = h\,, \overline{(M)}  = \bar M\,, \overline{(\mu)} = \mu\,, \overline{(D)} = D\,, \overline{(L^{ik})} = \varepsilon_{il} \varepsilon_{kj} L^{lj}\,, \nn
&& \overline{(\phi_i)} = \bar\phi^i\,, \quad\overline{(\sigma_i)} = \bar\sigma^i\,. \lb{RealHcomp}
\eea

\subsection{Superfield QK actions}
The universal $1D$ superfield action describing ${\cal N}=4$ supersymmetric QK mechanics
looks very simple \cite{Ivanov:2017ajf}
\bea
&&S_{QK} = \frac18 \int \mu_H \Big[ E\, \Big(\gamma q^{+ a}q^{-}_a - \hat{Q}^{+ r} \hat{Q}^{-}_{ r}\Big) + \beta \sqrt{E} \Big],  \quad \beta := - \gamma\frac{2}{\hat{\kappa}^2}\,, \label{SupLagr}
\eea
where $\mu_H = dtdw d^2\theta^+ d^2\theta^-$ is the measure of integration over the whole ${\cal N}=4, 1D$ superspace and
$q^{-a} :=D^{--}q^{+ a}\,, \,\hat{Q}^{-r} :=D^{--}\hat{Q}^{+ r}$. The specificity of the given model is concentrated in the harmonic constraints \p{qgen}, \p{Qgen}.
One or another choice of the QK potential ${\cal L}^{+4}$ generates one or another QK sigma model in the bosonic sector of \eqref{SupLagr}.

An important object appearing in \eqref{SupLagr} is the
harmonic-independent (in the central basis) supervielbein $E$ incorporating fields of some non-minimal version of ${\cal N}=4, 1D$ ``supergravity''.
Its presence secures invariance of \eqref{SupLagr}
under the appropriate local extension of ${\cal N}=4, 1D$ supersymmetry. The precise form of the superspace realization of this local supersymmetry
is given in \cite{Ivanov:2017ajf}, the realization on the component fields will be presented in the next Section.
With the choice $\gamma = \pm 1\,$, the action \eqref{SupLagr}
is invariant, respectively, under the extended global $Sp(n,1)$ $(\gamma =1)$ or $Sp(n+1)$ $(\gamma = -1)$ groups realized as
\be
\delta q^{+ a} = -\gamma \Lambda^{ar} \hat{Q}^+_r\,, \quad \delta \hat{Q}^{+ r} =  \Lambda^{ar} {q}^+_a\,, \label{ConstSp}
\ee
where $\Lambda^{ar}$ are the coset $Sp(n,1)/[Sp(1) \times Sp(n)]$  or $Sp(n+1)/[Sp(1) \times Sp(n)]$ constant parameters. The $Sp(1)$ and $Sp(n)$ subgroups in both cases are
realized as symplectic rotations with respect to the indices $a$ and $r$, respectively. Though the action \eqref{SupLagr} is formally invariant under \p{ConstSp} at any choice of
the QK potential ${\cal L}^{+4}$ in the harmonic constraints \p{qgen}, \p{Qgen}, the constraints themselves are covariant only with ${\cal L}^{+4} =0$, {\it i.e.} when they take
the linear form \p{HSconstrLin}. In this case for $\gamma = \pm 1$ the action \eqref{SupLagr} in the bosonic sector, after
fixing some gauges with respect to local symmetries,
describes $1D$ sigma models on the $\mathbb{H}{\rm H}^n$ and $\mathbb{H}{\rm P}^n$ target spaces (see Subsection 3.2). Thus the action \eqref{SupLagr}
with the constraints \p{HSconstrLin} describes the ``maximally symmetric'' particular
$\mathbb{H}{\rm H}^n$ or $\mathbb{H}{\rm P}^n$ cases of ${\cal N}=4$ supersymmetric QK mechanics \footnote{QK mechanics corresponding to any other homogeneous QK manifolds
are associated with some non-zero QK potentials in \p{qgen}, \p{Qgen}. A wide list of such QK potentials was given in ref. \cite{GIOO}.},
\bea
S_{HP} = \frac18 \int \mu_H \Big[ E\, \Big(\gamma q^{+ a}q^{-}_a - \hat{Q}^{+ r} \hat{Q}^{-}_{ r}\Big) + \beta \sqrt{E} \Big], \quad D^{++}q^{+a} = D^{++}\hat{Q}^{+k} = 0\,.
\label{SupLagrHP}
\eea

\section{Component QK Lagrangians}
In accord with the consideration in Sect. 2, three classes of the superfields which appear in \eqref{SupLagrHP} have the following field contents:
\vspace{0.2cm}

\begin{itemize}
\item The vielbein $E$ encompasses the multiplet ${\bf 8} + {\bf 8}$ of  the ``non-minimal'' ${\cal N}=4, 1D$ ``supergravity''.
Its bosonic sector
consists of the dynamical field $h(t)$ (``graviton'')  and the auxiliary fields $M(t), {\bar M}(t), \mu(t), D(t), L^{(ik)}(t)$.
The fermionic sector involves the dynamical fermionic fields $\phi_i(t), {\bar \phi}_i(t),$ (``gravitino'') and the auxiliary fields
$\sigma^i(t), {\bar \sigma}^i(t)$. The conjugation rules for the involved fields are given in \eqref{RealHcomp}.

\item Two other superfields, $q^{+a}$ and $\hat{Q}^{+ r}$,  describe $1D$ ``matter'' and both encompass the ${\cal N}=4, 1D$ multiplets $({\bf 4, 4, 0})$.
First of them contains the dynamical fields  $f^{ia}, \chi^a, {\bar \chi}^a$. It is $1D$ analog of the ``conformal compensator'' $q^+$ superfield of the harmonic superspace
formulation of ${\cal N}=2, 4D$ supergravity \cite{SGN2}, \cite{HSS}.
The set  of $({\bf 4, 4, 0})$ fields entering $\hat{Q}^{+ r}$ contains
the dynamical fields ${\hat F}^{ia}, \chi^r, {\bar \chi}^r\,$.  The conjugation properties of the relevant fields are given in \eqref{qQreal}.
\end{itemize}
The precise way how all these fields enter the respective superfields was already presented in \eqref{qcomp}, \eqref{Qcomp}.\\

We will be interested in the component Lagrangian $\mathbb{L}_{HP}$ which corresponds to the superfield one \eqref{SupLagrHP} and is obtained
from the latter after integrating there over Grassmann and harmonic variables. It is a sum of the three Lagrangians:

\begin{itemize}
\item The gauge-covariantized kinetic terms of the bosonic compensator and matter fields:
\bea
\mathbb{L}_{HP}^b &=& \frac12 h\,\Big(\dot{\hat{F}}^{ir}
\dot{\hat{F}}_{ir} - \gamma\,\dot{f}^{ia} \dot{f}_{ia}\Big) +
L_{ik}\Big(\hat{F}^{(i r} \dot{\hat{F}}^{k)}_{ r} - \gamma f^{(i a}
\dot{f}^{k)}_{ a}\Big) \nn && +\, \frac1{4}\,D\Big(\gamma
f^{ia}f_{ia} - \hat{F}^{i r}\hat{F}_{i r}
+\frac{\beta}{\sqrt{h}}\Big) \nn
&&
+\,\frac{\beta}{4}\,\frac{1}{\sqrt{h}h}\Big[L^{ik}L_{ik} -
\frac1{8}\big(M\bar M + \mu^2 +\dot{h}^2\big)\Big].\lb{Lagrbos}
\eea
\vspace{0.3cm}

\item The gauge covariantization of the kinetic terms of the fermionic compensator and matter fields:
\bea
\mathbb{L}^{f(1)}_{HP} &=& \frac{i}{4}h \left[\gamma\Big(\chi^a
\dot{\bar{\chi}}_a - \dot\chi^a {\bar{\chi}}_a\Big) - \chi^r
\dot{\bar{\chi}}_r + \dot\chi^r {\bar{\chi}}_r\right] \nn && +\,
\frac{i}{2}\phi_i\left(\gamma\dot{f}^{ia}\bar{\chi}_a-
\dot{\hat{F}}^{ir}\bar{\chi}_r\right) -
\frac{i}{2}\bar\phi_i\left(\gamma \dot{f}^{ia}{\chi}_a-
\dot{\hat{F}}^{ir}{\chi}_r\right) \nn
&&
+\,\frac{i}{2}\sigma_i\left(\gamma {f}^{ia}{\bar\chi}_a-
{\hat{F}}^{ir}{\bar\chi}_r\right) -
\frac{i}{2}\bar\sigma_i\left(\gamma {f}^{ia}{\chi}_a-
{\hat{F}}^{ir}{\chi}_r\right) \nn
&& +\,
\frac{M}{8}\Big(\gamma{\bar\chi}^a{\bar\chi}_a -
{\bar\chi}^r{\bar\chi}_r\Big) - \frac{\bar
M}{8}\Big(\gamma{\chi}^a{\chi}_a - {\chi}^r{\chi}_r\Big) \nn
&& +\,
\frac{\mu}{4}\Big(\gamma {\bar\chi}^a{\chi}_a - {\bar\chi}^r{\chi}_r\Big).
\lb{Ferm1}
\eea
\vspace{0.3cm}

\item The remaining Lagrangian that involves fermionic fields of the $1D$ ``supergravity'' multiplet and comes solely from the last term in \eqref{SupLagr}:
\bea
\mathbb{L}^{f(2)}_{HP} &=& \beta\, \frac{i}{32 h^{3/2}}
\Big(\phi^i\dot{\bar\phi}_i - \bar\phi^i\dot\phi_i + 4\sigma ^i
\bar\phi_i - 4\bar\sigma^i\phi_i\Big) \nn && +\, \beta\,\frac{3}{64
h^{5/2}}\Big(4iL^{ik}\phi_{(i}\bar\phi_{k)} + \frac{M}{2}
\bar\phi^i\bar\phi_i - \frac{\bar M}{2} \phi^i\phi_i + \mu
\phi^i\bar\phi_i \Big)\nn && +\, \beta\, \frac{15}{64\cdot
8}\,\frac{1}{h^{7/2}} (\phi^k\phi_k) (\bar\phi^i\bar\phi_i).
\lb{Ferm22}
\eea
\end{itemize}
\vspace{0.3cm}

The total off-shell Lagrangian is the sum of these three ones:

\bea
\mathbb{L}_{HP} = \mathbb{L}^{b}_{HP} + \mathbb{L}^{f(1)}_{HP} + \mathbb{L}^{f(2)}_{HP}\,. \lb{Total}
\eea


\subsection{Transformation properties of  QK Lagrangian}

The above Lagrangian is invariant under the local transformations with the parameters $b(t), \lambda^i(t), \tau^i_k(t)$ associated, respectively,
with the time reparametrizations, local ${\cal N}=4, 1D$ supersymmetry and local $SU(2)$ $R$-symmetry. Various sets
of fields have the following transformation laws:\\

\noindent{\it 1. The fields $f^{ia}, \chi^a, {\bar \chi}^a$}:
\vspace{0.1cm}

\bea
&& \delta_b f^{ia} = -2b\, \dot{f}^{ia} - \dot b\, f^{ia}\,, \; \delta_b\chi^a = -2b \dot{\chi}^a - 2\dot b\, \chi^a\,,  \;
\delta_b\,\bar{\chi}^a = -2b\, \dot{\bar\chi}^a - 2\dot b\, \bar{\chi}^a, \lb{qb} \\
&& \delta_\lambda f^{ia} = -\lambda^i\,\chi^a + \bar\lambda^i\,\bar{\chi}^a\,, \quad \delta_\lambda\chi^a = 2i\partial_t(\bar{\lambda}^i f_i^a)\,,
\;\delta_\lambda\bar\chi^a  = 2i\partial_t(\lambda^i f_i^a)\,, \lb{lambdaq}\\
&& \delta_\tau f^{ia} = \tau^i_{\;k}\,f^{ka}\,, \quad \delta_\tau \chi^a = \delta_\tau \bar{\chi}^a = 0\,. \lb{tauq}
\eea
\vspace{0.3cm}

\noindent{\it 2. The fields ${\hat F}^{ir}, \chi^r, {\bar \chi}^r$}:
\vspace{0.1cm}

\bea
&& \delta_b \hat{F}^{ir} = -2b \,\dot{\hat{F}}^{ir} - \dot b \,\hat{F}^{ir}\,, \; \delta_b\chi^r = -2b\, \dot{\chi}^r - 2\dot b\, \chi^r\,,  \;
\delta_b\bar\chi^r = -2b\, \dot{\bar\chi}^r - 2\dot b\, \bar{\chi}^r, \lb{Qb} \\
&& \delta_\lambda \hat{F}^{ir} = -\lambda^i\,\chi^r + \bar\lambda^i\,\bar{\chi}^r\,, \quad \delta_\lambda\chi^r = 2i\partial_t(\bar{\lambda}^i {\hat{F}}_i^r)\,,
\;\delta_\lambda\bar{\chi}^r  = 2i\partial_t(\lambda^i {\hat{F}}_i^r)\,, \lb{lambdaQ}\\
&& \delta_\tau  \hat{F}^{ir} = \tau^i_{\;k}\, \hat{F}^{kr}\,, \quad \delta_\tau \chi^r = \delta_\tau \bar{\chi}^r = 0\,. \lb{tauQ}
\eea
\vspace{0.3cm}

\noindent{\it 3. ${\cal N}=4, 1D$ ``supergravity'' fields}:
\vspace{0.1cm}

\bea
&&\delta_b h = -2b\,\dot{h} + 4\dot{b}\,h\,, \quad \delta_b M = -2b\,\dot{M} + 2\dot{b}\,M\,, \quad
\delta_b \mu = -2b\,\dot{\mu} + 2\dot{b}\,\mu\,, \nn
&&\delta_b L^{(ik)} = -2b\,\dot{L}^{(ik)} + 2\dot{b}\,L^{(ik)}\,, \quad
\delta_b D = -2b\,\dot{D} + 2\partial_t(\ddot{b}h)\,, \nn
&& \delta_b \phi^i = -2b\,\dot{\phi}^i + 3\dot{b}\,\phi^i\,,
\quad \delta_b \sigma^i = -2b\,\dot{\sigma}^i + \dot{b}\,\sigma^i +  \ddot{b}\phi^i\,, \lb{bH}
\eea
\bea
&& \delta_\lambda h = \lambda^i \phi_i - \bar\lambda^i\bar\phi_i\,, \;
\delta_\lambda  M = 2i\dot{\bar\lambda}^i \phi_i + i{\bar\lambda}^i (4\sigma_i - 2\dot{\phi}_i)\,, \nn
&& \delta_\lambda  \bar M = -2i\dot{\lambda}^i \bar{\phi}_i
- i{\lambda}^i (4\bar{\sigma}_i - 2\dot{\bar{\phi}}_i)\,, \nn
&& \delta_\lambda\mu = -i(\dot{\lambda}^i\phi_i +\dot{\bar\lambda}^i\bar\phi_i)
-{i}[\lambda^i(2\sigma_i - \dot{\phi}_i) + \bar\lambda^i(2\bar\sigma_i - \dot{\bar\phi}_i)] \nonumber \\
&& \delta_\lambda L^{(ik)} = \bar\lambda^{(i}\bar\sigma^{k)} - \lambda^{(i}\sigma^{k)}
-[\dot{\bar\lambda}^{(i}\bar\phi^{k)} - \dot{\lambda}^{(i}\phi^{k)}]\,, \nonumber \\
&& \delta_\lambda D = \lambda^i\dot{\sigma}_i - \bar\lambda^i\dot{\bar\sigma}_i
- \dot{\lambda}^i {\sigma}_i + \dot{\bar\lambda}^i{\bar\sigma}_i + \partial_t (\dot\lambda^i{\phi}_i
- \dot{\bar\lambda}^i{\bar\phi}_i)\,, \nn
&& \delta_\lambda \phi^i = \lambda^i M + \bar\lambda^i(\mu + i\dot{h}) + 4i \bar\lambda^k L^i_{\;k} - 4i\dot{\bar\lambda}^i h\,, \nn
&& \delta_\lambda \bar\phi^i = \bar\lambda^i \bar M - \lambda^i(\mu - i\dot{h}) + 4i \lambda^k L^i_{\;k} - 4i\dot{\lambda}^i h\,, \nn
&& \delta_\lambda \sigma^i = \dot{\lambda}^i M +
\dot{\bar\lambda}^i (\mu - i\dot{h}) + 2i \bar{\lambda}^k\dot{L}^i_{\;k} + i {\bar\lambda}^i D - 2i \ddot{\bar\lambda}^i h\,, \nn
&&\delta_\lambda {\bar\sigma}^i =  \dot{\bar\lambda}^i \bar M -
\dot{\lambda}^i (\mu + i\dot{h}) + 2i {\lambda}^k\dot{L}^i_{\;k} + i {\lambda}^i D - 2i \ddot{\lambda}^i h\,,
\lb{lambdaH}\\
&&{} \nn
&& \delta_\tau h = \delta_\tau M = \delta_\tau \mu = 0\,, \; \delta_\tau L^{(ik)} = h \dot{\tau}^{(ik)} - 2\tau^{(i m}L^{k)}_{\;m}\,, \;
\delta_\tau D = -2\dot{\tau}^{(ik)}L_{(ik)}\,, \nn
&& \delta_\tau \phi^i = \tau^i_{\;k}\phi^k\,, \quad \delta \sigma^i = \tau^i_{\;k}\sigma^k - \dot{\tau}^{(i k)}\phi_k\,. \lb{tauH}
\eea
\vspace{0.2cm}

The standard rigid ${\cal N}=4, 1D$ supersymmetry and $R$-symmetry $SU(2)$ transformations of the component fields
are recovered upon choosing the constant parameters in \eqref{qb} - \eqref{tauH}, $\dot\lambda^i = \dot b = \dot\tau^{ik} = 0\,.$
The ${\dot b} = 0$ transformations are just the constant time shifts.

The internal symmetry transformations \eqref{ConstSp} uniformly act in the evident way on the indices $a$ and $r$ of matter bosonic and fermionic fields,
properly mixing $f^{ia}$ with $\hat{F}^{ir}$ and $\chi^a$ with $\chi^r$.

It is not so easy to check that the Lagrangian $\mathbb{L}_{HP}$ given by eq. \p{Total} is indeed invariant, up to a total time derivative, under
all these sets of the field transformations. Nevertheless, this can be done, in complete agreement with the corresponding invariances
of the superfield action \eqref{SupLagr} proved in \cite{Ivanov:2017ajf}.

\subsection{Passing to the explicit $\mathbb{H}{\rm H}^n$ or $\mathbb{H}{\rm P}^n$ sigma model metrics}

Though in what follows we will deal with the off-shell ``ungauged'' Lagrangians \p{Lagrbos} - \p{Ferm22}, it is instructive to see how the standard
$\mathbb{H}{\rm H}^n$ or $\mathbb{H}{\rm P}^n$ sigma model Lagrangians are recovered from \p{Lagrbos}.

First, we omit the auxiliary fields by their algebraic equations of motion, $M = \bar M = \mu = 0$ (up to some pure fermionic terms). After this we rescale the fields as
\bea
f^{ia} = h^{-1/4} \tilde{f}^{ia}\,, \quad \hat{F}^{ir} = h^{-1/4} \tilde{F}^{ir}\,. \label{resc10}
\eea
Then, varying the auxiliary field $D$ yields the constraint
\bea
\gamma \tilde{f}^2 - \tilde{F}^2 - \gamma \frac{2}{\hat{\kappa}^2} = 0\,. \label{constr10}
\eea
The next step is to gauge away from $\tilde{f}^{ia}$ the symmetric part in the doublet indices using the local $SU(2)$ symmetry \p{tauq}:
\bea
\tilde{f}^{ia}(t) \;\Rightarrow \; \sqrt{2} \varepsilon^{ai} \,\tilde{\omega}(t)\,.
\eea
After substituting this into \p{constr10} we solve the latter for $\tilde{\omega}$ as
\bea
\tilde{\omega} = \frac{1}{\sqrt{2}|\hat\kappa|} \sqrt{ 1 + \gamma \frac{\hat{\kappa}^2}{2} \tilde{F}^2}\,. \label{omega10}
\eea
Finally, we eliminate the auxiliary field $L^{ik}$ by its algebraic equation of motion
\bea
L^{ik} = h\gamma {\kappa}^2\,\tilde{F}^{(i r}\, \dot{\tilde{F}}^{k)}_r\,. \label{LtildeF}
\eea

Plugging all these expressions into \p{Lagrbos}, we obtain
\bea
\mathbb{L}_{HP}^b = \frac12h^{1/2}\Big[\dot{\tilde{F}}^{ir}\dot{\tilde{F}}_{ir}
+ \gamma \hat{\kappa}^2 \dot{\tilde{F}}^{(ir}\tilde{F}^{k)}_r\dot{\tilde{F}}^{r}_{(i} \tilde{F}_{k)r} - \gamma \frac{\hat{\kappa}^2}{4}
\frac{1}{1 + \gamma\frac{\hat{\kappa}^2}{2}\, \tilde{F}^2}\,(\tilde{F}\dot{\tilde{F}}) (\tilde{F}\dot{\tilde{F}})\Big]. \label{BosFin}
\eea
Note that the term $\sim (\dot{h})^2$ drops out from the final Lagrangian. The Lagrangian takes a simpler form after redefining
\bea
&&\tilde{F}^{ir} = \frac{F^{ir}}{\sqrt{1 - \gamma \frac{\hat{\kappa}^2}{2}\, {F}^2}}\,, \lb{NewPar} \\
&&\mathbb{L}_{HP}^b = \frac12h^{1/2}\Big[\frac{1}{1 - \gamma \frac{\hat{\kappa}^2}{2}\, {F}^2} (\dot{F}\dot{F})
+ \gamma \frac{\hat{\kappa}^2}{(1 - \gamma \frac{\hat{\kappa}^2}{2}\, {F}^2)^2 }\,(F^{ir}F_i^s)\,(\dot{F}^{k}_r\dot{F}_{ks})\Big],  \lb{NewPar2}
\eea
whence one can read off the target metric as
\bea
&& \mathbb{L}_{HP}^b  =\frac12h^{1/2} \, g_{ir\; ks} \dot{F}^{ir} \dot{F}^{ks}\,,\lb{Metr1} \\
&& g_{ir\; ks} = \varepsilon_{ik}\,\frac{1}{1 - \gamma \frac{\hat{\kappa}^2}{2}\, {F}^2}\Big[ \Omega_{rs}
+ \gamma \hat{\kappa}^2\frac{1}{1 - \gamma \frac{\hat{\kappa}^2}{2}\, {F}^2} F^j_r F_{js} \Big]. \lb{Metr2}
\eea
The nonlinearly realized coset $Sp(1,n)/[Sp(1)\times Sp(n)]$ $(\gamma = 1)$ and $Sp(1+n)/[Sp(1)\times Sp(n)]$ $(\gamma = -1)$ transformations leaving
invariant \p{Metr1} read
\bea
\delta F^{ir} = \Lambda^{ir} - \gamma\hat{\kappa}^2 \Lambda^{sk} F^r_{k} F_s^i\,.
\eea

Note that the Lagrangians \p{BosFin} and \p{NewPar2} reveal an obvious analogy with the Lagrangians of the massless ``relativistic particle'' on the aforementioned QK spaces
$\mathbb{H}{\rm H}^n$ and $\mathbb{H}{\rm P}^n$: in the gauge $h =1$ they become the corresponding $1D$  sigma model Lagrangians, while varying
with respect to the ``einbein'' $h$ yields a sort of the particle $P^2 \simeq 0$ constraint on these $n$-dimensional manifolds (in the Hamiltonian approach, this is
just vanishing of the relevant Hamiltonian, see Sect. 5).

\section{Equations of motion}
It is straightforward to derive the classical equations of motion following from the action with the Lagrangian $\mathbb{L}_{HP}$, eq. \p{Total}.
They are divided into sets of non-dynamical algebraic equations, as well as the dynamical equations, of the second order in $\partial_t$ for bosonic fields and
of the first order for fermionic fields.\\

\noindent{\it 1. Non-dynamical equations:}
\vspace{0.2cm}

\be
D: \ \ \gamma f^2 - \hat{F}^2 + \frac{\beta}{h^{1/2}} = 0\,,\lb{ConstrBase}
\ee

\bea
L^{ik} : \ \ && L^{ik} = -2 \frac{h^{3/2}}{\beta} \Big[\hat{F}^{(i r}\dot{\hat{F}}^{k)}_r - \gamma {f}^{(i a}\dot{f}^{k)}_a \Big]
+ \frac{3i}{8 h} \bar\phi^{(i}\phi^{k)}\,, \lb{L} \\
{\bar M}: \ \ && M = \frac{4 h^{3/2}}{\beta} \Big(\chi^r\chi_r - \gamma \,\chi^a \chi_a\Big) -\frac{3}{4 h}\,\phi^i\phi_i\,, \nn
M: \ \ && \bar M =\frac{4 h^{3/2}}{\beta} \Big(\gamma \,\bar\chi^a \bar\chi_a - \bar\chi^r\bar\chi_r\Big) +\frac{3}{4 h}\,\bar\phi^i\bar\phi_i\,,\nn
\mu: \ \ && \mu =\frac{4 h^{3/2}}{\beta} \Big(\gamma \,\bar\chi^a \chi_a - \bar\chi^r \chi_r\Big) + \frac{3}{4 h}\, \phi^i \bar\phi_i\,, \lb{MbarMmu}
\eea

\bea
{\bar \sigma}_i: \ \ \phi^i = \frac{4 h^{3/2}}{\beta } \Big(\gamma f^{ia}\chi_a - \hat{F}^{i r}\chi_r \Big), \quad
\sigma_i: \ \ \bar\phi^i = \frac{4 h^{3/2}}{\beta } \Big(\gamma f^{ia}\bar\chi_a - \hat{F}^{i r}\bar\chi_r \Big).\lb{phiConstr}
\eea
\vspace{0.1cm}

\noindent{\it 2. Dynamical equations:}
\vspace{0.2cm}

\bea
{\hat F}^{ir}: \ \ && \partial_t(h \dot{\hat{F}}_{ir}) + \frac12 D \hat{F}_{ir} - 2L_{ik}\dot{\hat{F}}_{r}^k  - \dot{L}_{ik}\hat{F}_{r}^k
-\frac{i}2 \big(\phi_i{\dot{\bar\chi}}_r - {\bar\phi}_i{\dot\chi}_r\big) \nn
&& +\,\frac{i}2 \big[(\sigma_i - \dot{\phi}_i)\bar{\chi}_r -  ({\bar\sigma}_i - {\dot{\bar\phi}}_i){\chi}_r\big]
 = 0
\lb{Feq}
\eea
(equation for $f^{ia}$ has the same form, with the evident substitution of indices $r \rightarrow a$).

\bea
h: \ \ \partial^2_t (h^{-\frac{1}{2}}) = \frac{4}{\beta}({\dot{\hat F}}^2 - \gamma{\dot f}^2) + V, \lb{heq}
\eea
\bea
&&V =  -\frac{D}{h^{3/2}} - \frac{3}{h^{5/2}}[L^{ik}L_{ik} -\frac{1}{8}(M{\bar M} +\mu^2 +{\dot h}^2)
+\frac{i}{8}(\phi^i{\dot {\bar \phi}}_i -{\bar \phi}^i{\dot \phi}_i + 4\sigma^i{\bar \phi}_i - 4{\bar \sigma}^i\phi_i)]  \nn
&&+\frac{2i}{\beta}\left[\gamma\Big(\chi^a{\dot{\bar \chi}} _a-{\dot \chi}^a{\bar \chi}_a\Big) - \chi^r{\dot{\bar \chi}} _r+{\dot \chi}^r{\bar \chi}_r\right]
- \frac{15}{16h^{7/2}}(4iL^{ik}\phi_i{\bar \phi}_k +\frac{M}{2}{\bar \phi}^i{\bar \phi}_i -\frac{{\bar M}}{2}\phi^i\phi_i  \nn
&&\quad \quad \quad \quad+ \mu\phi^i{\bar \phi}_i) -\frac {7\times 15}{128 h^{9/2}}\phi^k\phi_k{\bar \phi}^i{\bar \phi}_i.
\eea

Then follow the equations for matter fermions:
\bea
{\bar \chi}^r:   && \dot{\chi}_r + \frac{1}{h}[-\phi_i \dot{\hat{F}}_{r}^i - \sigma_i  {\hat{F}}_{r}^i +\frac{i}2 M  {\bar\chi}_r
+ \frac{i}2\mu \chi_r +\frac{1}{2 }\dot h \chi_r] = 0\,, \nn
{\chi}^r:&& \dot{\bar\chi}_r + \frac{1}{h}[ - {\bar\phi}_i \dot{\hat{F}}_{r}^i - {\bar\sigma}_i {\hat{F}}_{r}^i +\frac{i}2 \bar{M}  {\chi}_r  - \frac{i}2\mu{\bar\chi}_r
+\frac12 \dot h {\bar\chi}_r] = 0 \lb{Chieq}
\eea
(and similar equations for $\chi_a$ and $\bar\chi_a$).\\

Finally, we obtain the equations for the ``gravitino'' $\phi^i$ and $\bar\phi^i$:
\bea
{\bar \phi}^i: &&{\dot{\phi}}_i -2 \sigma_i +\frac{3i}4 h^{-1}\big[4 i L_i^k\phi_k + M{\bar\phi}_i + (\mu + i\dot h)\phi_i\big] -
\frac{8}{\beta} h^{3/2} \big(\gamma\dot{f}^a_i\chi_a - \dot{\hat{F}}^r_i \chi_r\big) \nn
&&+\, \frac{15i}{16}h^{-2}{\bar\phi}_i (\phi^k\phi_k) = 0\,, \nn
\phi^i:&&{\dot{\bar\phi}}_i -2 {\bar\sigma}_i -\frac{3i}4 h^{-1}\big[-4 i L_i^k{\bar\phi}_k - \bar M{\phi}_i + (\mu - i\dot h){\bar\phi}_i\big]
- \frac{8}{\beta} h^{3/2} \big(\gamma\dot{f}^a_i{\bar\chi}_a - {\dot{\hat{F}}}^r_i {\bar\chi}_r\big) \nn
&&-\,\frac{15i}{16}h^{-2}{\phi}_i ({\bar\phi}^k{\bar\phi}_k) = 0\,.\lb{Phieq}
\eea

\section{Noether charges}
Noether charges are calculated in the standard way for the transformations with constant parameters.
\vspace{0.2cm}

\noindent {\bf 1. R-symmetry}. We start with the $SU(2)$ current associated with the parameters $\tau_{ik}$. We define
$$\delta_\tau \Phi_A \frac{\partial \mathbb{L}}{\partial \dot{\Phi}_A} = \tau^{ik} J_{ik}$$
whence

\bea
&& J_{ik} =h \big[ \hat{F}_{(i r} \dot{\hat{F}}^r_{k)} -\gamma f_{(i a} \dot{f}^a_{k)}\big] +
\frac{i}2 \big[ \phi_{(i} \big(\hat{F}_{k)}^r \bar\chi_r - \gamma f^a_{k)} \bar\chi_a \big) - \bar\phi_{(i}\big(\hat{F}_{k)}^r \chi_r - \gamma f^a_{k)}\chi_a \big)\big]\nn
&& -\,\frac12 \big(\hat{F}^2 - \gamma f^2\big) L_{ik} + i\beta \frac{1}{16 h^{3'2}}\,\phi_{(i}\bar\phi_{k)}. \lb{Jik}
\eea
Using the bosonic constraint \p{ConstrBase} and the on-shell expression for $L_{ik}$ and $\phi_i, \bar\phi_i$ (from the non-dynamical equations \p{L} and \p{phiConstr}),
it is easy to show that on shell
\bea
J_{ik} = 0\,. \lb{Jonshell}
\eea
\vspace{0.3cm}

\noindent {\bf 2. Supersymmetry}. Next we construct supercharges associated with the constant parameters $\lambda^i$.  We define
\bea
\delta_\lambda \mathbb{L}_{HP} = \partial_t U\,, \quad \delta_\lambda\Phi_A \frac{\partial \mathbb{L}_{HP}}{\partial \dot{\Phi}_A} - U
= \lambda^k Q_k + \bar\lambda^k \bar{Q}_k\,, \lb{defQ}
\eea
with
\bea
&&U = \lambda^k\{\frac{h}{2}({\dot{\hat{F}}}_k^r\chi_r - \gamma{\dot f}^a_k\chi_a)+L^i_k({\hat F}^r_i\chi_r -\gamma f_i^a\chi_a) -\frac{1}{4} \sigma_k({\hat F}^2 -\gamma f^2) \nn
&&-\, \frac{i}{4}\phi_k(\chi^r{\bar \chi}_r - \gamma\chi^a{\bar \chi}_a)+\frac{i}{4}{\bar \phi}_k(\chi^r{ \chi}_r - \gamma \chi^a{\chi}_a) \nn
&&+\,\frac{\beta}{4h^{1/2}}[\sigma_k-\frac{i}{8h}(M{\bar \phi}_k +\mu\phi_k - i{\dot h}\phi_k +4iL^i_k\phi_i)- \frac{3i}{32h^2}{\bar \phi}_k\phi^i\phi_i]\} \nn
&&+\, {\bar \lambda}^k\{-\frac{h}{2}({\dot{\hat{F}}}{}^r_k {\bar \chi}_r - \gamma {\dot f}^a_k{\bar \chi}_a)-L^i_k({\hat F}^r_i{\bar \chi}_r -\gamma f_i^a{\bar \chi}_a)
+\frac{1}{4} {\bar \sigma}_k({\hat F}^2 - \gamma f^2) \nn
&&-\,\frac{i}{4}{\bar \phi}_k(\chi^r{\bar \chi}_r - \gamma \chi^a{\bar \chi}_a)+\frac{i}{4}{\phi}_k({\bar \chi}^r{ \bar \chi}_r - \gamma {\bar \chi^a}{\bar \chi}_a)  \nn
&&+\,\frac{\beta}{4h^{1/2}}[-{\bar \sigma}_k+\frac{i}{8h}({\bar M}{ \phi}_k -\mu{\bar \phi}_k - i{\dot h}{\bar \phi}_k +4iL^i_k{\bar \phi}_i)
- \frac{3i}{32h^2}{ \phi}_k{\bar \phi}^i{\bar \phi}_i]\},
\eea
whence
\bea
&& Q_k = h\big(\dot{\hat{F}}^r_k \chi_r - \gamma\dot{f}^a_k \chi_a\big) + \frac14 \sigma_k \big({\hat{F}}^{jr} {\hat{F}}_{jr} - \gamma {f}^{ak}f_{ak}\big) \nn
&&-\, \frac{i}{4} \phi_k \big( \chi^r\bar{\chi}_r - \gamma\chi^a\bar{\chi}_a\big) + \frac{i}{4} \bar\phi_k \big( \chi^r{\chi}_r - \gamma \chi^a{\chi}_a\big) \nn
&& +\, \frac{\beta}{16 h^{3/2}}\big[-4h\sigma_k -\dot{h}\phi_k - 4 L_k^l\phi_l + \frac{3i}{8 h} \bar\phi_k(\phi^l\phi_l)\big], \lb{Q}\\
&& \bar{Q}_k = h\big(\dot{\hat{F}}^r_k \bar{\chi}_r -\gamma\dot{f}^a_k \bar{\chi}_a\big) + \frac14 \bar{\sigma}_k \big({\hat{F}}^{jr} {\hat{F}}_{jr} - \gamma {f}^{ak}f_{ak}\big) \nn
&&+\, \frac{i}{4} \bar\phi_k \big( \chi^r\bar{\chi}_r - \gamma \chi^a\bar{\chi}_a\big) -\frac{i}{4} \phi_k \big( \bar{\chi}^r\bar{\chi}_r - \gamma \bar\chi^a\bar{\chi}_a\big) \nn
&& -\, \frac{\beta}{16 h^{3/2}}\big[4h\bar\sigma_k +\dot{h}\bar\phi_k + 4 L_k^l\bar\phi_l + \frac{3i}{8 h} \phi_k(\bar\phi^l\bar\phi_l)\big]. \lb{barQ}
\eea
\vspace{0.3cm}

Let us show that $Q_i = 0$ on shell. We identically rewrite the expression in the first term in \p{Q} as
\bea
&& \dot{\hat{F}}^r_k \chi_r -\gamma \dot{f}^a_k \chi_a = \partial_t\big({\hat{F}}^r_k \chi_r - \gamma {f}^a_k \chi_a \big) - \big({\hat{F}}^r_k \dot\chi_r - \gamma {f}^a_k \dot\chi_a) \nn
&& =\, -\beta\big( \frac{1}{4 h^{3/2}} \dot\phi_k -\frac{3}{8}\frac{1}{h^{5/2}} \dot{h}\phi_i \big)  -\frac{1}{h}\phi_j \big( \hat{F}^r_k \dot{\hat{F}}^j_r
- \gamma {f}^a_k \dot{{f}}^j_a\big) \nn
&& +\, \frac{\beta}{2 h^{3/2}} \sigma _k -i\frac{\beta}{8 h^{5/2}}\big(M\bar\phi_k + \mu \phi_k + i\dot{h} \phi_k\big), \lb{Check1}
\eea
where we made use of the bosonic constraint \p{ConstrBase}, eq. \p{Chieq}, together with the analogous one for $\chi_a$  and (a few times)
of the fermionic constraint \p{phiConstr}. As the next steps we represent
$$
-\frac{1}{h}\phi_j\big(\hat{F}^r_k \dot{\hat{F}}^j_r - \gamma{f}^a_k \dot{{f}}^j_a\big) = -\frac{\beta}{8 h^{5/2}} \dot{h} \phi_k
+ \frac{1}{h}\phi^j\big[\hat{F}^r_{(k} \dot{\hat{F}}_{j)r} - \gamma {f}^a_{(k} \dot{{f}}_{j)a} \big],
$$
substitute it in \p{Check1} and then use the equation of motion for $\phi_i$, eq. \p{Phieq}. After some algebra, we obtain
\bea
&& \dot{\hat{F}}^r_k \chi_r -\gamma \dot{f}^a_k \chi_a  = \frac{\beta}{16 h^{5/2}}\dot{h}\phi_k - \frac{1}{h}\phi^j\big[\hat{F}^r_{(k} \dot{\hat{F}}_{j)r}
- \gamma{f}^a_{(k} \dot{{f}}_{j)a} \big] \nn
&& -\,i\frac{\beta}{16 h^{5/2}}\big[ M\bar\phi_k + \mu \phi_k + 12i L_k^l\phi_l + \frac{15}{4 h}\bar\phi_k(\phi^l\phi_l)\big]\,. \lb{Check2}
\eea
Finally, using \p{L}, we obtain
\bea
&& \dot{\hat{F}}^r_k \chi_r -\gamma\dot{f}^a_k \chi_a  = \frac{\beta}{16 h^{5/2}}\dot{h}\phi_k  \nn
&& -\,i\frac{\beta}{16 h^{5/2}}\big[ M\bar\phi_k + \mu \phi_k + 4i L_k^l\phi_l + \frac{3}{2 h}\bar\phi_k(\phi^l\phi_l)\big],  \lb{Check22}
\eea
and this expression exactly cancels the remaining terms in \p{Q}, taking into account that the total coefficient of $\sigma_i$ in \p{Q} is vanishing as a consequence
of the constraint \p{ConstrBase}.

In a similar fashion or just by conjugation we obtain:
\bea
&& \dot{\hat{F}}^r_k {\bar \chi}_r - \gamma\dot{f}^a_k {\bar \chi}_a  = \frac{\beta}{16 h^{5/2}}\dot{h}{\bar \phi}_k  \nn
&& -\,i\frac{\beta}{16 h^{5/2}}\big[ {\bar M}\phi_k - \mu {\bar \phi}_k + 4i L_k^l{\bar \phi}_l - \frac{3}{2 h}\phi_k(\phi^l\phi_l)\big],  \lb{Check2'}
\eea
Thus on shell
\bea
Q_k = \bar{Q}_k = 0\,. \lb{Q0}
\eea

It is worth mentioning that, using the  identities \p{Check22},\p{Check2'} deduced above, the equations for the auxiliary fields $\sigma_i ,\,{\bar\sigma}_i$
\p{Phieq} can be simplified:

\bea
&&{\dot{\phi}}_i -2 \sigma_i +\frac{i}{4h}\big[4 i L_i^k\phi_k + M{\bar\phi}_i + (\mu + i\dot h)\phi_i\big] +
 \frac{3i}{16}h^{-2}{\bar\phi}_i (\phi^k\phi_k) = 0\,, \nn
&&{\dot{\bar\phi}}_i -2 {\bar\sigma}_i -\frac{i}{4 h}\big[-4 i L_i^k{\bar\phi}_k - \bar M{\phi}_i + (\mu - i\dot h){\bar\phi}_i\big]
-\frac{3i}{16}h^{-2}{\phi}_i ({\bar\phi}^k{\bar\phi}_k) = 0\,.\lb{Phieq'}
\eea
\vspace{0.3cm}

\noindent {\bf 3. Time translations}. We define the conserved charge associated with $b$ transformations (Hamiltonian) as
\bea
\delta_b\mathbb{L}_{HP} = -2b \dot{\mathbb{L}}_{HP}\,, \quad \dot\Phi_A \frac{\partial \mathbb{L}_{HP}}{\partial \dot\Phi_A} - \mathbb{L}_{HP} = \mathbb{H}\,, \lb{defH}
\eea
whence
\bea
 \mathbb{H} =  \mathbb{H}^b +  \mathbb{H}^{f(1)} + \mathbb{H}^{f(2)}\,, \nonumber
\eea
where
\bea
\mathbb{H}^b &=& \frac12 h\,\Big(\dot{\hat{F}}^{ir}
\dot{\hat{F}}_{ir} - \gamma\,\dot{f}^{ia} \dot{f}_{ia}\Big)
-  \frac1{4}\,D\Big(\gamma
f^{ia}f_{ia} - \hat{F}^{i r}\hat{F}_{i r}
+\frac{\beta}{\sqrt{h}}\Big) \nn
&&
-\,\frac{\beta}{4}\,\frac{1}{h^{3/2}}\Big[L^{ik}L_{ik} -
\frac1{8}\big(M\bar M + \mu^2 -\dot{h}^2\big)\Big],\lb{Hbos}
\eea

\bea
 \mathbb{H}^{f(1)} &=& -\frac{i}{2}\sigma_i\left(\gamma{f}^{ia}{\bar\chi}_a-
{\hat{F}}^{ir}{\bar\chi}_r\right) +
\frac{i}{2}\bar\sigma_i\left(\gamma {f}^{ia}{\chi}_a-
{\hat{F}}^{ir}{\chi}_r\right) \nn
&& -\,
\frac{M}{8}\Big(\gamma {\bar\chi}^a{\bar\chi}_a -
{\bar\chi}^r{\bar\chi}_r\Big) + \frac{\bar
M}{8}\Big(\gamma {\chi}^a{\chi}_a - {\chi}^r{\chi}_r\Big) \nn
&&-\, \frac{\mu}{4}\Big(\gamma{\bar\chi}^a{\chi}_a - {\bar\chi}^r{\chi}_r\Big),
\lb{HFerm1}
\eea

\bea
\mathbb{H}^{f(2)} &= &  -  \beta\,\frac{3}{64
h^{5/2}}\Big(4iL^{ik}\phi_{(i}\bar\phi_{k)} + \frac{M}{2}
\bar\phi^i\bar\phi_i - \frac{\bar M}{2} \phi^i\phi_i + \mu
\phi^i\bar\phi_i \Big)\nn
&&-\beta\, \frac{i}{8 h^{3/2}}
\Big(\sigma ^i
\bar\phi_i  - \bar\sigma^i\phi_i\Big) - \beta\, \frac{15}{64\cdot
8}\,\frac{1}{h^{7/2}} (\phi^k\phi_k) (\bar\phi^i\bar\phi_i).
\lb{Ferm2}
\eea
Putting these formulas together and using some equations for the auxiliary fields we get:
\bea
&&\mathbb{H} = \frac12 h\,\Big(\dot{\hat{F}}^{ir}
\dot{\hat{F}}_{ir} - \gamma\,\dot{f}^{ia} \dot{f}_{ia}\Big) -\,\frac{\beta}{4h^{3/2}}\,\Big[L^{ik}L_{ik} - \frac1{8}\big(M\bar M + \mu^2 -\dot{h}^2\big)\Big]\nn
&& \quad \quad \qquad \,-\frac{3i\beta}{16
h^{5/2}}L^{ik}\phi_{i}\bar\phi_{k}-  \frac{15\beta}{64\cdot
8 h^{7/2}}\,(\phi^k\phi_k) (\bar\phi^i\bar\phi_i)\,. \lb{Hami}
\eea
It still remains to show that
\be
 \mathbb{H} = 0 \lb{H0}
\ee
on shell.

The proof of \p{H0} is more involved compared to \p{Q0}. The basic step is to represent the first term in the first line of \p{Hbos} as
\bea
 \frac{h}{2} \,\Big(\dot{\hat{F}}^{ir}\dot{\hat{F}}_{ir} - \gamma\,\dot{f}^{ia} \dot{f}_{ia}\Big) = \frac14 \partial_t\big[h \partial_t\big(\hat{F}^2 - \gamma f^2\big)\big]
 -\frac12 \big[\hat{F}^{ir}\partial_t\big(h {\dot {\hat F}}_{ir}\big) - \gamma {f}^{ia}\partial_t\big(h {\dot {f}}_{ia}\big)\big].
\eea
Then, making use of the equations of motion, we obtain:
\bea
 \frac12 h\,\Big(\dot{\hat{F}}^{ir}\dot{\hat{F}}_{ir} - \gamma\,\dot{f}^{ia} \dot{f}_{ia}\Big) = -\frac{\beta}{4} \partial_t^2(h^{1/2})  +W, \lb{FFeq}
 \eea
 where
 \bea
&&W = \frac{1}{2}{\hat F}^{ir}\Big\{\frac{1}{2}D{\hat F}_{ir} - 2L_{ik}{\dot{\hat F}}^k_r -{\dot L}_{ik}{{\hat F}}^k_r
-\frac{i}{2}(\phi_i{\dot{\bar \chi}}_r -{{\bar \phi}}{\dot \chi}_r) \nn
&& +\,\frac{i}{2}[(\sigma_i - {\dot \phi}_i){\bar \chi}_r-({\bar \sigma}_i - {\dot {\bar \phi}}_i){\chi}_r]\Big\} -
\gamma \frac{1}{2}{f}^{ia}\Big\{\frac{1}{2}D{f}_{ia} - 2L_{ik}{\dot{f}}^k_a -{\dot L}_{ik}{{f}}^k_a \nn
&&-\, \frac{i}{2}(\phi_i{\dot{\bar \chi}}_a
-{{\bar \phi}}{\dot \chi}_a) + \frac{i}{2}[(\sigma_i - {\dot \phi}_i){\bar \chi}_a-({\bar \sigma}_i - {\dot {\bar \phi}}_i){\chi}_a]\Big\}.
\eea
As the next step, we cast \p{heq} in the form:
\bea
-\frac{1}{h}\partial^2_t (h^{1/2}) +\frac{{\dot h}^2}{2h^{5/2}}=\frac{4}{\beta}({\dot{\hat F}}^2 - \gamma {\dot f}^2) + V\,.
\eea
Using this equation, we rewrite \p{FFeq} as the following identity:
\bea
 \frac{h}{2}\,\Big(\dot{\hat{F}}^{ir}\dot{\hat{F}}_{ir} - \gamma\,\dot{f}^{ia} \dot{f}_{ia}\Big) =  -\frac{\beta h}{4}V +\frac{\beta}{8h^{3/2}}{\dot h}^2 - W\,.
 \eea

 Now, the strategy will be following: the terms which contain $\dot h $ are going to eventually cancel, the other terms with time derivatives, or matter fermions, can be
 replaced by using the equations of motion. In this way we end up with an
 expression for $\mathbb{H}$ which contains only terms related to the supergravity multiplet and its auxiliary fields ordered by odd integer powers of
 $h^{-\frac{1}{2}}$, and the coefficients of each individual power of this sort  must  vanish. This can be rather easily checked for the terms with
 $h^{-\frac{1}{2}}$ and  $h^{-\frac{3}{2}}$. More involved calculation shows that all terms with the higher inverse degrees of $h$ also vanish.

Note that the vanishing of the (super)charges associated with the rigid world-line symmetries is in accordance with the $1D$ version of the second Noether
theorem (see \cite{Avery}, \cite{PKT} for a modern discussion of this theorem and its implications). Moreover, since ${\cal N}=4$ supersymmetry
combine these (super)charges into a supermultiplet, it would be enough to show the on-shell vanishing only for one of them, say for $Q_i, \bar Q^i$. Nevertheless, the
above explicit checks provide a good verification of the self-consistency of our approach. To be firmly confident of the correctness of the expressions for Noether
(super)charges is important for the construction of the Hamiltonian formalism (actually, its purely bosonic truncation) in the next sections.
\vspace{0.3cm}

\noindent{\bf 4. $Sp(n,1)$ and $Sp(n+1)$ symmetries}. Finally, we will discuss the rigid isometries. The general definition of the relevant conserved current is
$$\delta_\Lambda \Phi_A \frac{\partial \mathbb{L}}{\partial \dot{\Phi}_A} = \Lambda^{ar} J_{ar}\,.$$
Then
\bea
J_{ar} &=&  f^i_a \Big[h \dot{\hat F}_{ir} -\frac{i}{2}\big(\phi_i\bar\chi_r - \bar\phi_i \chi_r\big)\Big] +
\hat{F}^i_r\Big[ h\dot{f}_{ia}-\frac{i}{2}\big(\phi_i\bar\chi_a - \bar\phi_i \chi_a\big)\Big] \nn
&&-\, 2L_{ik}f^i_a \hat{F}^k_r + \frac{i}{2}h \Big(\chi_r\bar\chi_a - \bar\chi_r\chi_a \Big)\,. \label{SpCurr}
\eea
The $Sp(1)$ and $Sp(n)$ currents are calculated analogously. All these currents are conserved, but non-vanishing on shell,
because they correspond to the internal global symmetries as opposed to the previously presented currents associated with the worldline
symmetries.

\section{Canonical momenta}
The canonical momenta are calculated straightforwardly:
\bea
&& {\cal P}_{ir}^{(F)} = \frac{\partial \mathbb{L}}{\partial \dot{\hat{F}}^{ir}} = h \dot{\hat{F}}_{ir} - L_{ik}\hat{F}_{r}^k -\frac{i}2 \big(\phi_i \bar{\chi}_r
- \bar\phi_i \chi_r\big), \nn
&& {\cal P}_{ia}^{(f)} = \frac{\partial \mathbb{L}}{\partial \dot{f}^{ia}} = -\gamma h  \dot{f}_{ia} + \gamma L_{ik}{f}_{a}^k +\frac{i}2 \gamma \big(\phi_i \bar{\chi}_a
- \bar\phi_i \chi_a\big), \nn
&& {\cal P}^{(h)} = \frac{\partial \mathbb{L}}{\partial \dot{h}} = - \frac{\beta}{16 h^{3/2}}\,\dot{h}\,, \lb{BosMom} \\
&&{\,} \nn
&&{\cal P}^{(\chi)}_r = \frac{\partial \mathbb{L}}{\partial \dot{\chi}^r} = -\frac{i}4 h \bar{\chi}_r\,, \quad
{\cal P}^{(\bar\chi)}_r = \frac{\partial \mathbb{L}}{\partial \dot{\bar\chi}^r} = \frac{i}4 h {\chi}_r\,, \nn
&&{\cal P}^{(\chi)}_a = \frac{\partial \mathbb{L}}{\partial \dot{\chi}^a} = \frac{i}4 \gamma h \bar{\chi}_a\,, \quad
{\cal P}^{(\bar\chi)}_a = \frac{\partial \mathbb{L}}{\partial \dot{\bar\chi}^a} = -\frac{i}4 \gamma  h {\chi}_a\,, \nn
&&{\cal P}^{(\phi)}_i = \frac{\partial \mathbb{L}}{\partial \dot{\phi}^i} =  i\beta \frac{1}{32 h^{3/2}}\, \bar\phi_i \,, \quad
{\cal P}^{(\bar\phi)}_i = \frac{\partial \mathbb{L}}{\partial \dot{\bar\phi}^i} = -i\beta \frac{1}{32 h^{3/2}}\, \phi_i \,, \lb{FermMom}
\eea
where the derivatives with respect to $\dot{\chi}^r, \dot{\bar \chi}^r, \dot{\chi}^a, \dot{\bar \chi}^a, \dot{\phi}^i$ and $\dot{\bar\phi}^i$ are understand as the right ones.

The supercharges can be expressed in terms of the bosonic canonical momenta as

\bea
Q_k &=& {\cal P}^{(F) r}_k \chi_r + {\cal P}^{(f) a}_k \chi_a + {\cal P}^{(h)} \phi_k  + \frac{3i\beta}{16\cdot 8 h^{5/2}} \,\bar\phi_k (\phi^l\phi_l) \nonumber \\
&& +\,\frac{i}4 \big[ \phi_k(\bar\chi^r\chi_r - \gamma \bar\chi^a \chi_a) -\bar\phi_k(\chi^r\chi_r - \gamma \chi^a \chi_a)\big], \lb{Q22} \\
\bar Q_k &=& {\cal P}^{(F) r}_k \bar\chi_r + {\cal P}^{(f) a}_k \bar\chi_a + {\cal P}^{(h)} \bar\phi_k  - \frac{3i\beta}{16\cdot 8 h^{5/2}} \,\phi_k (\bar\phi^l\bar\phi_l) \nonumber \\
&& +\,\frac{i}4 \big[ \phi_k(\bar\chi^r\bar\chi_r - \gamma \bar\chi^a \bar\chi_a) -\bar\phi_k(\bar\chi^r\chi_r - \gamma \bar\chi^a \chi_a)\big]. \lb{barQ22}
\eea
We used here the algebraic equations \p{ConstrBase} and \p{phiConstr}. Analogously, one can reexpress the $SU(2)$ current and the Hamiltonian
\bea
J_{kl} &=& {\cal P}^{(F) r}_{(k} \hat{F}_{l)r} + {\cal P}^{(f) a}_{(k} f_{l)a} + \frac{i\beta}{16 h^{3/2}}\, \phi_{(k}\bar\phi_{l)}\,, \lb{J22} \\
\mathbb{H} &=& \frac{1}{2h}\Big[ ({\cal P}^{(F)})^2 - \gamma ({\cal P}^{(f)})^2 -\frac{16 h^{5/2}}{\beta} ({\cal P}^{(h)})^2 \Big]  \nonumber \\
&& -\, \frac{i\beta}{4 h^{5/2}}\, L^{(ik)} \phi_{(i}\bar\phi_{k)}  - \frac{\beta}{32 h^{3/2}}\, \big(M\bar M + \mu^2\big)
- \frac{3\beta}{8\cdot 64 h^{7/2}} (\phi^i\phi_i) (\bar\phi^i\bar\phi_i) \nonumber \\
&& + \, \frac1{h} L^{ik}J_{ik} + \frac{i}{2h} \big(\phi_i \bar Q^i - \bar\phi_i Q^i \big). \label{H22}
\eea
Since $J_{ik} = Q^i = \bar Q^i =0$ on shell, the last line in the expression \p{H22} can be suppressed. The auxiliary fields $M, \bar M$ and $\mu$ can be replaced
by their on-shell expressions.

Finally, the internal symmetry current $J_{ar}$, \eqref{SpCurr}, is expressed as
\bea
J_{ar} = f^i_a {\cal P}^{(F)}_{i r} - \gamma \hat{F}^i_r {\cal P}^{(f)}_{ia} +\frac{i}{2}h \big(\chi_r\bar{\chi}_a - \bar{\chi}_r\chi_a\big).
\eea

Now one can define the Poisson brackets, quantize them and find quantum expressions for the (super)charges.
The wave function $|\Phi>$ should satisfy the conditions
\bea
J_{ik}|\Phi> = Q^i |\Phi>= \bar Q^i|\Phi> = \mathbb{H}|\Phi> =0
\eea
and, after solving these equations,  be expressed in terms of irreps of $Sp(n+1)$ (for $\gamma = - 1$) or $Sp(1, n)$ (for $\gamma = 1$).

Due to the first-class constraints $J_{kl} \simeq  0\, \; Q^i \simeq 0\,, \; \bar{Q}_i \simeq 0\,, \;\mathbb{H} \simeq 0$ the considered system, prior to quantization, should be
exposed to an accurate Hamiltonian analysis, which for the bosonic sector will be performed in the next section.  We will finish this section by presenting the set of Poisson
brackets for the dynamical variables.
\vspace{0.2cm}

\noindent{\it Bosonic brackets}.
\bea
\{\hat{F}^{ir}, {\cal P}^{(F)}_{k s}\} = \delta^i_k\delta^r_s\,, \quad \{{f}^{ia}, {\cal P}^{(f)}_{k b}\} = \delta^i_k\delta^a_b\,, \quad \{ {h}, {\cal P}^{(h)}\} = 1\,. \lb{BosBr}
\eea
\vspace{0.2cm}

\noindent{\it Fermionic  brackets}.
\vspace{0.1cm}

To unambiguously define the brackets involving fermionic fields we need to apply to the Dirac method. As is seen from the expressions for the fermionic momenta \eqref{FermMom},
there is a set of second-class constraints
\bea
&& \varphi_r^{(\chi)} := {\cal P}^{(\chi)}_r + \frac{i}4 h \bar{\chi}_r \simeq 0\,, \quad \bar{\varphi}_r^{(\chi)} := {\cal P}^{(\bar{\chi})}_r - \frac{i}4 h {\chi}_r \simeq 0\,, \nn
&& \varphi_a^{(\chi)} := {\cal P}^{(\chi)}_a  - \frac{i}4\,\gamma\, h \bar{\chi}_a \simeq 0\,, \quad \bar{\varphi}_a^{(\chi)} := {\cal P}^{(\bar{\chi})}_a
+ \frac{i}4\,\gamma\, h {\chi}_a \simeq 0\,, \nn
&& \varphi_i^{(\phi)} := {\cal P}^{(\phi)}_i -i \beta \frac{1}{32 h^{3/2}}\,\bar\phi_i \simeq 0\,, \quad \bar{\varphi}_i^{(\phi)}
:= {\cal P}^{(\bar\phi)}_i +i \beta \frac{1}{32 h^{3/2}}\,\phi_i
\simeq 0\, \label{2Class}
\eea
with the following non-zero canonical brackets:
\bea
\{\varphi_r^{(\chi)}, \bar{\varphi}_s^{(\chi)}\} = \frac{i}{2} h \Omega_{rs}\,, \;
\{\varphi_a^{(\chi)}, \bar{\varphi}_b^{(\chi)}\} = -\frac{i}{2}\,\gamma\, h \varepsilon_{ab}\,, \;
\{\varphi_i^{(\phi)}, \bar{\varphi}_k^{(\phi)}\} = - i\beta \frac{1}{16 h^{3/2}}\,\varepsilon_{ik}\,. \label{Dirak1}
\eea

Then the standard Dirac procedure yields the following non-zero brackets involving the fermionic variables:
\bea
&&\{\chi_r, \bar\chi_s \}_{D} = 2 i \frac{1}{h}\,\Omega_{rs}\,, \, \{\chi_a, \bar\chi_b \}_{D} = -2 i\gamma \frac{1}{h}\,\varepsilon_{ab}\,, \,
\{\phi_i, \bar{\phi}_k \}_{D} = -16 i\, \frac{h^{3/2}}{\beta}\, \varepsilon_{ik}, \label{PurFerm} \\
&& \{{\cal P}^{(h)}, \chi_r \}_{D} = \frac{1}{2h} \chi_r\,, \quad \{{\cal P}^{(h)}, \bar{\chi}_r \}_{D} = \frac{1}{2h} \bar{\chi}_r\,, \nn
&& \{{\cal P}^{(h)}, \chi_a \}_{D} = \frac{1}{2h} \chi_a\,, \quad  \{{\cal P}^{(h)}, \bar{\chi}_a \}_{D} = \frac{1}{2h} \bar{\chi}_a\,, \nn
&&\{{\cal P}^{(h)}, \phi_i \}_{D} = -\frac{3}{4 h}\,\phi_i\,, \quad \{{\cal P}^{(h)}, \bar{\phi}_i \}_{D} = -\frac{3}{4 h}\,\bar{\phi}_i\,. \label{PhFerm}
\eea
In what follows, we will omit the index ``$D$'' on these brackets. Using them, one can calculate the brackets between $J_{kl}\,, \, Q^i\,, \,  \bar{Q}_i\,,$ and
$\mathbb{H}$ and convince oneself that they form a closed superalgebra. Here we present the brackets between the $Sp(n,1) (Sp(n+1))$ currents $J_{ar}$:
\bea
\{J_{ar}, J_{b s}\} = -2\gamma \Big[\,\Omega_{rs}\, J_{(ab)} + \varepsilon_{ab}\, J_{(rs)}\,\Big], \lb{SpComm}
\eea
where
\bea
J_{(ab)} = f^i_{(a}\,{\cal P}^{(f)}_{i b)} -\frac{i}{2}\gamma\,h\,\chi_{(a}\bar{\chi}_{b)}\,, \quad J_{(rs)} = \hat{F}^i_{(r}\,{\cal P}^{(F)}_{i s)}
+ \frac{i}{2} h\,\chi_{(r}\bar{\chi}_{s)} \lb{SpCurr2}
\eea
are $Sp(1)$ and $Sp(n)$ currents. Note that the brackets of ${\cal P}^{(h)}$ with the internal symmetry currents $J_{ar}\,, \;J_{(ab)}$ and $J_{(rs)}$ are vanishing,
as it should be. The same concerns the brackets with the Hamiltonian $\mathbb{H}$. It is rather easy to check that $J_{ar}\,, \;J_{(ab)}$ and $J_{(rs)}$ have zero bracket
with the quadratic combination of the currents
\bea
C^{(2)} = \gamma \,J^{bs} J_{bs} - 2 \Big[ J^{(ab)}J_{(ab)} + J^{(rs)}J_{(rs)}\Big],  \lb{Casimir}
\eea
which can thus be identified with the second-order Casimir of $Sp(n, 1)$ (for $\gamma =1$) or $Sp(n+1)$ (for $\gamma = -1\,$).

Actually, besides the dynamical variables, the full ungauged (super)charges involve
the auxiliary fields, bosonic and fermionic, which have no kinetic terms in the Lagrangian and so have vanishing conjugate momenta. This  produces new Hamiltonian constraints,
which, for the time being, are difficult to analyze in a full generality\footnote{The Hamiltonian analysis of the simplest ${\cal N}=1, 1D$ ``supergravity'' system,
with the worldline multiplet consisting of a real ``graviton'' and one real ``gravitino'', was accomplished in \cite{VanHolten}.}. For this reason, in this paper we limit
our consideration to the bosonic sector of the whole system.

Our goal is to weigh the way covariant quantization works. We did this by determining the global charges corresponding to the
relevant symmetries, which vanish as a consequence of the equations of motion, as a short way to generate some of the
constraints which will appear in the full constrained Hamiltonian approach. Subsequently we perform the full analysis of the
bosonic sector which shows that our approach is correct.

\section{Dirac analysis of the bosonic model}

In what follows we will restrict our attention to the bosonic part  \p{Lagrbos}:

\bea
\mathbb{L}_{HP}^b &=& \frac12 h\,\Big(\dot{\hat{F}}^{ir}
\dot{\hat{F}}_{ir} - \gamma\,\dot{f}^{ia} \dot{f}_{ia}\Big) +
L_{ik}\Big(\hat{F}^{(i r} \dot{\hat{F}}^{k)}_{ r} - \gamma f^{(i a}
\dot{f}^{k)}_{ a}\Big) \nn && +\, \frac1{4}\,D\Big(\gamma
f^{ia}f_{ia} - \hat{F}^{i r}\hat{F}_{i r}
+\frac{\beta}{\sqrt{h}}\Big) \nn
&&
+\,\frac{\beta}{4}\,\frac{1}{\sqrt{h}h}\Big(L^{ik}L_{ik} -
\frac1{8}\dot{h}^2\Big),\lb{Lagrbos'}
\eea
where we ignored the fields $M, {\bar M}$ and $\mu$
which fully decouple in the absence of fermionic variables.  We will perform the Dirac analysis of the above Lagrangian.

We prefer to deal with the ``ungauged'' Lagrangian \eqref{Lagrbos'} instead of the nonlinear  sigma model Lagrangian  \eqref{Metr1}, \eqref{Metr2},
because the former involves  all the original fields (including the auxiliary ones) and the corresponding unfixed gauge invariances
suggested by the original harmonic superspace formulation. We hope that this approach, being less nonlinear, will translate into a more acceptable
quantum version, with less ordering problems. Moreover, it is rather non-trivial task to solve the constraints introduced by the field $D(t)$ for the more involved
cases with non-vanishing QK potential,  ${\cal L}^{+ 4} \neq 0$, including all inhomogeneous QK cases \cite{Ivanov:1999vg}. We hope that our ``ungauged''
approach could help to avoid (or at least relax) this problem as well. Anyway, to the best of our knowledge,  nobody considered the Hamiltonian approach and the relevant quantization
either for \eqref{Lagrbos'} or for \eqref{Metr1}, \eqref{Metr2} before.
\vspace{0.2cm}

\noindent{\it Digression: relativistic particle}.
\vspace{0.1cm}

As a warmup, we replay the Dirac formalism in the application to the well known case of the massive relativistic particle:
\bea
L= -m\left({ -({\dot {\mathbb{X}}})^2}\right)^{1/2},\quad ({\dot {\mathbb{X}}})^2 = -({\dot {\mathbb{X}}}^0)^2 + ({\dot {\vec {\mathbb{X}}}})^2. \lb{origL}
\eea
We have:
\bea
\mathbb{P} = \frac{\partial L}{\partial {\dot {\mathbb{X}}}} = \frac{m {\dot {\mathbb{X}}}}{\sqrt{ -{\dot {\mathbb{X}}}^2}}, \Rightarrow \ \mathbb{P}^2
= -(P_0)^2 + {\vec P}^2=\frac{m^2 ({\dot {\mathbb{X}}})^2}{{ -{\dot {\mathbb{X}}}^2}} = -m^2,
\eea
that is, we arrive at the primary constraint:
\bea
\Phi(t) = \mathbb{P}^2 + m^2 \simeq 0.\lb{RP1}
\eea
The corresponding canonical Hamiltonian vanishes and so \p{RP1} is the only first-class constraint. The Hamiltonian becomes:
\bea
\mathbb{H } = \frac{1}{2}\ell\left(\mathbb{P}^2 +m^2\right).
\eea
We consider now the Hamiltonian form of the Lagrangian for the relativistic massive particle:
\bea
L  = {\dot {\mathbb{X}}} \mathbb{P} - \frac{\ell}{2}( \mathbb{P}^2 + m^2).\lb{massivepart}
\eea
This Lagrangian has a local
symmetry:
\bea
\delta \mathbb{X} = \alpha (t) \mathbb{P},\quad \delta \mathbb{P} = 0, \quad \delta \ell = {\dot \alpha}(t),
\eea
for which reason we naturally expect first-class constraints. Indeed, $\mathbb{X}$ and $\mathbb{P}$ are conjugate variables, while the momentum
corresponding to the variable $\ell$ vanishes:
\be
\Pi_\ell = \frac {\partial L}{\partial {\dot \ell}} = 0 \Rightarrow \ \varPhi_\ell = \Pi_\ell \simeq 0,
\ee
so that  $ \varPhi_\ell  \simeq 0$ is a primary constraint. Introducing the Hamiltonian:
\be
\mathbb{H} = {\dot {\mathbb{X}}}\mathbb{P} - L=  \frac{\ell}{2}( \mathbb{P}^2 + m^2),
\ee
together with the relevant non-vanishing equal-time Poisson brackets,
\be
\left\{ \mathbb{X}, \mathbb{P} \right\} = 1, \quad \left\{ \ell, \Pi_\ell \right\} = 1,
\ee
and then imposing the condition of conservation of the constraint $\varPhi_\ell$,
\be
{\dot {\varPhi_\ell}} = \left\{\varPhi_\ell, \mathbb{H} \right\} = -\frac{1}{2}( \mathbb{P}^2 + m^2) \simeq 0,
\ee
we arrive at the mass-shell condition $( \mathbb{P}^2 + m^2) \simeq 0$ as a secondary first-class constraint. This is the complete set of constraints,
and the quantum theory is obtained by replacing the non-vanishing Poisson brackets by the commutators:
\be
\left[ \mathbb{X}, \mathbb{P} \right] = i, \quad \left[ \ell, \Pi_\ell \right] = i\,.
\ee
Thus $\mathbb{X}, \ell$ can be viewed as standard multiplication operators, while the corresponding momenta
are $\mathbb{P} = \partial /( i \partial \mathbb{X}),\ \Pi_\ell = \partial/ (i\partial \ell)$. These operators
act on the space of wave functions $\Psi(\mathbb{X}, \ell)$, and the corresponding first-class constraints tell us that
the wave function does not depend on $\ell$ and obeys the Klein Gordon equation:
\be
( \mathbb{P}^2 + m^2)\Psi(\mathbb{X}) = 0.
\ee
Getting back to \p{massivepart}, we can eliminate $\mathbb{P}$:
\bea
{\dot {\mathbb{X}}} =  \ell\mathbb{P},\quad \Rightarrow \quad L =\frac{
 {\dot {\mathbb{X}}}^2}{2\ell} - \frac{\ell m^2}{2} \lb{momentum}
 \eea
and, furthermore, eliminate $\ell$:
 \bea
 \frac{
 -{\dot {\mathbb{X}}}^2}{2\ell^2} - \frac{ m^2}{2} =0, \ \Rightarrow \  L =  -m\left({ -({\dot {\mathbb{X}}})^2}\right)^{1/2}.
 \eea
Then it becomes clear that the Lagrangian in the Hamiltonian form \p{massivepart} is more general then \p{origL},
as it also permits the case $m=0$. For the massless particle the configuration space action involves an additional field $\ell$. The auxiliary field $\ell$ is necessary to  guarantee the gauge invariance:
\bea
\delta \mathbb{X} = \frac{\alpha (t)}{\ell}{\dot{ \mathbb{X}}},\quad  \quad \delta \ell = {\dot \alpha}(t).
\eea

This auxiliary field cannot be eliminated as this would lead to a vanishing action. Its role is to impose the gauge invariant mass shell constraint
(this is nicely explained  in  \cite{Green:1987sp}). One can then fix the gauge to obtain a simpler action. In what follows we
will proceed for  \p{Lagrbos'}  in a similar fashion.
\vspace{0.2cm}

\noindent{\it Back to \p{Lagrbos'}}.
\vspace{0.1cm}

A first glance at \p{Lagrbos'} tells us that it is not in the Hamiltonian form analogous to \p{massivepart}.
Exploiting the Dirac method, we will arrive at the corresponding Hamiltonian form.

We start by defining the canonical momenta:
$$
 {\cal P}_{ir}^{(F)} = \frac{\partial \mathbb{L}_{HP}}{\partial \dot{\hat{F}}^{ir}} = h \dot{\hat{F}}_{ir} - L_{ik}\hat{F}_{r}^k , \quad
 {\cal P}_{ia}^{(f)} = \frac{\partial \mathbb{L}_{HP}}{\partial \dot{f}^{ia}} = -\gamma h  \dot{f}_{ia} + \gamma L_{ik}{f}_{a}^k , \nn
$$\bea
 {\cal P}^{(h)} = \frac{\partial \mathbb{L}_{HP}}{\partial \dot{h}} = - \frac{\beta}{16 h^{3/2}}\,\dot{h}\,, \ \
 {\cal P}^{(L)}_{ik} = \frac{\partial \mathbb{L}_{HP}}{\partial {\dot{L}}^{ik}} = 0, \quad
{\cal P}^{)(D)} = \frac{\partial \mathbb{L}}{\partial{\dot{D}}} = 0\,.
 \lb{FermMom'}
\eea
An inspection of the above formulas tells us that we have now two primary sets of commuting constraints:
\bea
{\Phi}^{(L)}_{ik} :=  {\cal P}^{(L)}_{ik} \simeq 0, \quad {\Phi}^{(D)} := {\cal P}^{{D}} \simeq 0.  \lb{Primary}
\eea
Indeed, using the non-vanishing Poisson brackets:
$$
\left\{{\hat{F}}^{ir},  {\cal P}_{js}^{(F)}\right\} = \delta_j^i\delta_s^r, \ \ \left\{{{f}}^{ia},  {\cal P}_{jb}^{(f)}\right\} = \delta_j^i\delta_b^a, \ \  \left\{D,  {\cal P}^{(D)}\right\} = 1,
$$
\bea
\left\{{{L}}^{ik},  {\cal P}_{jl}^{(L)}\right\} = \frac{1}{2}\left(\delta_j^i\delta_l^k + \delta_l^i\delta_j^k \right), \quad \mbox{and} \quad   \left\{h,  {\cal P}^{(h)}\right\} = 1,
\eea
we can calculate:
\bea
\left\{{\Phi}^{(L)}_{ik},  {\Phi}^{(D)}\right\} = 0.
\eea
Introducing the Hamiltonian of our system as
$$
\mathbb{H}^\prime =  \dot{\hat{F}}^{ir} {\cal P}_{ir}^{(F)} + \dot{{f}}^{ia} {\cal P}_{ia}^{(f)}
+  {\dot h}  {\cal P}^{(h)} + [(  {\dot{L}}^{ik} {\cal P}^{(L)}_{ik} + {\dot D}{\cal P}^{)(D)}) = 0] - \mathbb{L}_{HP}^b ,
$$
we obtain:
$$
\mathbb{H}^\prime = \frac{1}{2h}\left(  {\cal P}_{ir}^{(F)} +L_{ik} {\hat {F}}^k_r\right)\left( {\cal P}^{(F)ir} -L^{ik^\prime}{\hat {F}}_{k^\prime}^r\right)
-  \frac{\gamma}{2h}\left( {\cal P}_{ia}^{(f)} - \gamma L_{ik} { {f}}^k_a\right)\left({\cal P}^{(f)ia} + \gamma L^{ik^\prime}{{f}}_{k^\prime}^a\right)
$$
\be
-\frac{8h^{3/2}}{\beta}({\cal P}^{(h)} )^2 -\frac{D}{4}\left(\gamma f^2 - {\hat F}^2 + \frac{\beta}{h^{3/2}}\right) - \frac{\beta}{4h^{3/2}}L^{ij}L_{ij}.\lb{Hamilt}
\ee
Imposing the requirement of weak vanishing of the Poisson bracket of the primary constraints \p{Primary} with the Hamiltonian \p{Hamilt},
we obtain two new sets of first-class constraints:
\bea
\Phi^{F-f} := \gamma f^2 - {\hat F}^2 +\frac{\beta}{h^{1/2}}  \simeq 0, \quad \Phi_{l l^\prime}^{F{\cal P^{(F)}}}
:={\hat F}_{(l}^r{\cal P}_{l^\prime)r}^{(F)} + { f}_{(l}^a{\cal P}_{l^\prime)a}^{(f)} \simeq 0. \lb{constraints}
\eea
Using the definitions \p{FermMom'} in \p{Hamilt} and taking into account the constraints themselves, one gets:
\be
\mathbb{H}^\prime   \simeq \mathbb{H}\left|_{\left(M,{\bar M}, \mu, \phi_i, {\bar \phi}_k\right) =0}\right.,
\ee
where $\mathbb{H}$ is given by \p {Hami}.
Using eq. \p {FermMom'}  in $\Phi_{l l^\prime}^{F{\cal P^{(F)}}}$ we get:
$$
\Phi_{i k}^{F{\cal P^{(F)}}} = -J_{ik}\left| _{\left({\phi_i, {\bar \phi}_j}\right)=0}\right. = h{\hat F}^r_{(i}{\dot {\hat{F}}}_{k)r} + \frac{1}{2} {\hat F}^2L_{ik},
$$
where $J_{ik} $  is given by eq. \p{Jik}. This means that the constraint $\Phi_{l l^\prime}^{F{\cal P^{(F)}}}$ is just the $SU(2)$ - $R$ charge discussed earlier.

Further, one calculates:
$$
\left\{\Phi^{F-f},\Phi_{l l^\prime}^{F{\cal P^{(F)}}} \right\} = 0,
$$
$$
\left\{\Phi_{l l^\prime}^{F{\cal P}^{(F)}} , \Phi_{k k^\prime}^{F{\cal P}^{(F)}} \right\} = \frac{1}{2}\left(\epsilon_{lk} \Phi_{l^\prime k^\prime}^{F{\cal P}^{(F)}}
+ \epsilon_{l^\prime k} \Phi_{l k^\prime}^{F{\cal P}^{(F)}} + \epsilon_{lk^\prime} \Phi_{l^\prime k}^{F{\cal P}^{(F)}}
+ \epsilon_{l^\prime k^\prime} \Phi_{l k}^{F{\cal P}^{(F)}}\right) .
$$
At this stage, we can use these constraints to simplify the Hamiltonian \p{Hamilt}:
\bea
\mathbb{H}^\prime
 &=&  \frac{1}{2h}\Big[  ( {\cal P}^{(F)})^2 -\frac{1}{\gamma} ( {\cal P}^{(f)})^2 -\frac{16h^{5/2}}{\beta}({\cal P}^{(h)} )^2\Big]
 -\frac{1}{4}\left(\frac{L^{ij}L_{ij}}{h} +D\right)\Phi^{F-f}\nn
 && -\, 2L^{ij}\Phi_{i j}^{F{\cal P^{(F)}}} \simeq H,
\eea
with
\bea
H=  \frac{1}{2h}\Big[ ( {\cal P}^{(F)})^2 -\frac{1}{\gamma} ( {\cal P}^{(f)})^2 -\frac{16h^{5/2}}{\beta}({\cal P}^{(h)} )^2\Big].\lb{redHamilt}
\eea
Then, $\Phi_{l l^\prime}^{F{\cal P^{(F)}}}$ commutes with $H$, while commuting $\Phi^{F-f} $ with $H$ gives rise to the new constraint:
\bea
\Phi ^{{\cal P}^{(f)}f} := \frac{1}{h}\left({\cal P}^{(f) i a}f_{ia} + {\cal P}^{(F) i r}{\hat F}_{ir}\right) - 4{\cal P}^{h}  \simeq 0, \lb{fdot}
\eea
Poisson bracket of which with $H$ defined in \p{redHamilt} is equal to:
$$
\left \{\Phi ^{{\cal P}^{(f)}f} , H\right\} \simeq  -\frac{2}{h}H.
$$
We also have:
$$
\left \{\Phi ^{{\cal P}^{(f)}f} , \Phi_{l l^\prime}^{F{\cal P^{(F)}}}\right\} =0,\quad\left \{\Phi ^{{\cal P}^{(f)}f} ,\Phi^{F-f}\right\} =-\frac{2}{h}\Phi^{F-f}.
$$
Taking into account that the new constraint $H$ is weakly equivalent to $\mathbb{H}^\prime$, which in its turn is weakly equivalent to
$\mathbb{H}\left|_{{M,{\bar M}, \mu, \phi_i, {\bar \phi}_k} =0} \right.$, the properly restricted \p {Hami} is weakly equivalent to the new constraint $H$.
It is thereby established  that, in the present approach, the generators of global symmetries of our model correspond to some first-class constraints.

 The algebra of the constraints $\big(\Phi^{F-f}, \Phi_{l l^\prime}^{F{\cal P^{(F)}}}, \Phi ^{{\cal P}^{(f)}f} , H\big)$, closes in the weak sense and the
Hamiltonian finally becomes:
$$
{\cal H} = H_0(t) H - A^{ll^\prime}(t)\Phi^{F{{\cal P}^{(F)}}}_{l l^\prime} + B(t)\Phi ^{{\cal P}^{(f)}f} - D(t)\Phi^{F-f}.
$$
Now we are able to write the action in the Hamiltonian form:
\bea
S=  \int_0^{t_1}dt\Big[\dot{\hat{F}}^{ir} (t){\cal P}_{ir}^{(F)}(t) + \dot{{f}}^{ia}(t) {\cal P}_{ia}^{(f)} (t)+  {\dot h}(t)  {\cal P}^{(h)} (t)- {\cal H}(t)\Big]. \lb{Haction}
\eea
Using the definitions \p{FermMom'} in \p{fdot}, one can show that
\bea
\Phi ^{{\cal P}^{(f)}f} =-\frac{1}{2}\frac{d}{dt}\Phi^{F-f}.
\eea
Therefore, integrating by parts in the action,  one can absorb the constraint $\Phi ^{{\cal P}^{(f)}f} $
into a redefinition of the Lagrange multiplier $D$. We will not further pursue this approach, because, when we deduce
the corresponding gauge transformations, some of them may turn out singular on the surface of constraints. The action \p {Haction}
should have the corresponding gauge invariances specified by the transformations which are generated by the first-class constraints incorporated in this action.
In what follows we will spell the local transformations which should not be singular on the surface of constraints.
\vspace{0.2cm}

\noindent{\bf A}. Transformations corresponding to the constraint $H$ \p{redHamilt}:
\bea
\delta {\hat F}^{ir} = \frac{b(t)}{h} {\cal P}^{(F)\ ir}, \ \delta { f}^{ia} = -\frac{b(t)}{\gamma h} {\cal P}^{(f)\ ia}, \
\delta h = -b\frac{16{h^{3/2}}}{\beta}{\cal P}^{(h)}, \       \lb{btransf}
\eea
$$\delta {\cal P}^{(h)} = \frac{b(t)}{h}H + \frac{20b(t)h^{1/2}}{\beta}({\cal P}^{(h)})^ 2,\ \delta H_0 = {\dot b} +\frac{2b}{h} B, \
\delta B = -2bD -\frac{16 b h^{1/2}}{\beta}{\cal P}^{(h)} B.$$
\vspace{0.1cm}

\noindent{\bf B}. Transformations corresponding to the constraint $\Phi_{l l^\prime}^{F{\cal P^{(F)}}}$ in \p{constraints}:
$$
\delta {\hat F}^{ir} =\tau^{il}{\hat F}^r_l, \ \delta {\cal P}^{F}_{ir} =\tau_i^l{\cal P}^{(F)}_{rl}, \  \delta { f}^{ia}
=\tau^{il}{f}^a_l, \  \delta {\cal P}^{(f)ia} ,=\tau_i^l{\cal P}^ (f)_{al},
$$
\bea
\delta A^{ll^\prime} = -{\dot \tau}^{l{l'}}- \tau^l_kA^{kl'} - \tau^{l'}_kA^{kl}.
\eea
\vspace{0.1cm}

\noindent{\bf C}. Next, transformations corresponding to the constraint $\Phi^{F-f} $ in \p{constraints}:
\bea
\delta {\cal P}^{F}_{ir} =2 \alpha{\hat F}_{ir}, \  \delta {\cal P}^{f}_{ia} =-2 \gamma\alpha{ f}_{ia}, \
\delta {\cal P}^{(h)} = \frac{\beta \alpha}{2h^{3/2}}, \lb{F-f}
\eea
\bea
\delta D = -{\dot \alpha} - \frac{2\alpha}{h}B, \ \delta B = 2\alpha H_0.
\eea
\vspace{0.1cm}
\noindent{\bf D}. Finally, transformations corresponding to the constraint $\Phi ^{{\cal P}^{(f)}f} $ in \p{fdot}:
$$
\delta {\hat F}^{ir} =\frac{c}{h}{\hat F}^{ir}, \ \delta {\cal P}^{(F)}_{ir} =
-\frac{c}{h}{\cal P}^{(F)}_{ir}, \  \delta { f}^{ia} =\frac{c}{h}{f}^{ia}, \  \delta {\cal P}^{(f)}_{ia}
= -\frac{c}{h}{\cal P}^ {(f)}_{ia}, \ \delta h = -4c,
$$
$$\delta{\cal P}^{(h)} = \frac{c}{h^2}(f^{ia}{\cal P}^{(f)} +{\hat F}^{ir}{\cal P}^{(F)}_{ir}),\ \delta H_0 = -\frac{2c}{h} H_0, \ \delta B
= {\dot c} +\frac{16h^{1/2}c}{\beta}{\cal P}^{(h)}H_0,\ \delta D = -\frac{2c}{h} D.
$$

 We thus have established that the action \p{Haction} is invariant under the local transformations listed above. We have shown that this action
 has six gauge invariances,
 while for the original action \p{Lagrbos'} we exhibited  only four explicit gauge invariances, {\it viz.}, the local $SU(2)$, and time reparametrizations.
 An interesting question is as to whether the action  \p{Lagrbos'} also exhibits the remaining two gauge invariances.
 However, it may happen  that such additional gauge invariances are specific just to the Hamiltonian form of the action.
 As in the case of the Hamiltonian form of the action for the massive particle, we expect the presence of the relevant first-class constraints. Indeed,
${\cal P}_{ir}^{(F)}(t), {\cal P}_{ia}^{(f)} (t), {\cal P}^{(h)} (t) $ are  variables canonically conjugated to ${\hat{F}}^{ir} (t),  {{f}}^{ia}(t),  { h}(t)$,
while the momenta corresponding to $H_0,   A^{ll^\prime}, B,  D,$ vanish, so that the standard Poisson brackets restore the previous four constraints
as secondary first-class ones. The quantization of the above system is problematic as at the moment we are not aware of the general solution
to the first-class constraints on the space of wave functions $\Psi({\hat F}^{ir}, f^{ia}, h)$.  Gauge fixing might simplify the above constraints,
but we are interested to glimpse the covariant quantization.

It remains to prove the equivalence with the original action. To this end,
like for the massive particle in \p{momentum}, we substitute the corresponding momenta in \p{Haction} by their explicit expressions:
$$
 {\cal P}^{{(F)}{ir}} = \frac{h}{H_0}\left( \dot{\hat{F}}^{ir} + A^{il}\hat{F}^{r}_l - \frac{B}{h}{\hat F}^{ri}\right) , \
 {\cal P}^{{(f)}{ia}} = -\frac{\gamma h}{H_0} \left( \dot{f}^{ia} + A^{il}{f}^{a}_l  - \frac{B}{h} f^{ia} \right), \
 $$

 $$
 {\cal P}^{(h)} = - \frac{\beta}{16H_0 h^{3/2}}\left({\dot h} + 4B\right)\, .\,
 $$
After the redefinition $L^{ij} = hA^{ij}$,  some rearrangements and integrations by parts we represent the Lagrangian in \p{Haction} as:
$$
L = \frac{1}{H_0}\left[\mathbb{L}_{HP}^b - \frac{D}{4}\Phi^{F-f}\right] + \left[\frac{4H_0 +1}{4H_0}D
-\frac12\frac{d}{dt}\left(\frac{B}{H_0}\right) - \frac{B^2}{2hH_0}-\frac{L^{ij}L_{ij}}{4hH_0}\right]\Phi^{F-f}.
$$
If we eliminate $H_0$ from the above expression we obtain that the Lagrangian is weakly vanishing.
This can be easily confirmed using the available constraints. Like in the case of the massless particle, it is a step which can be done but
it does not appear to be useful. To obtain our original action it is in fact enough to fix the gauge with respect to the transformations \p{btransf}
by the condition $H_0 =1$.
In order to be convinced that this gauge choice is permissible, we assume that $H_0 = 1+ \epsilon$  and, using \p{btransf},
determine the infinitesimal transformation parameter to be:
$$
b = \left(\int^t dt"\epsilon(t")exp \left\{2 \int^{t"}du\frac{B}{h}(u)\right\}  \right) exp \left\{-2\int^t dt'\frac{B}{h}(t')\right\}.
$$
Finally, in the gauge $H_0 = 1$  we redefine the auxiliary field $D$ as
$$
\frac {D'}{4} = \frac{5D}{4} -\frac12\frac{dB}{dt} - \frac{B^2}{2h} - \frac{L^{ij}L_{ij}}{4h}
$$
and come back to the initial Lagrangian \p{Lagrbos'}. We therefore conclude that the action \p{Haction} and the action corresponding to \p{Lagrbos'} are equivalent.

\section{Concluding remarks}
In this paper we continued the study of the new class of ${\cal N}=4$ supersymmetric mechanics models introduced in \cite{Ivanov:2017ajf},
the Quaternion-K\"ahler (QK) ones.
We limited our attention to their simplest representatives, with the $1D$ sigma models on the homogeneous projective manifolds $\mathbb{H}{\rm H}^n$ or
$\mathbb{H}{\rm P}^n$ as the bosonic core. We started from the total off-shell component actions of these supersymmetric models, wrote down
the local gauge transformations leaving these
actions invariant, and explicitly presented the corresponding global invariances, together with the Noether currents associated with the latter.
The full set of the equations of motion for different fields, involving both dynamical and algebraic equations, was accurately written down.
The currents corresponding to global symmetries the gauging of which yields the total local symmetries of the action, were found to be vanishing
on the shell of the equations of motion, while those related to the global isometries $Sp(n+1)$ (or $Sp(n,1)$) do not vanish. The vanishing of the first type of currents
is quite analogous of the on-shell vanishing of Virasoro currents in bosonic string theory and/or the vanishing of ${\cal N}=4$ supercurrents in the spinning particle
coupled to a non-propagating ${\cal N}=4, 1D$ supergravity (see, e.g., \cite{PaSor}) \footnote{An important difference from
the spinning particle is that the models of ${\cal N}=4$ QK mechanics
are coupled to an extended (``non-minimal'') worldline ${\cal N}=4, 1D$ supergravity multiplet involving, besides the gauge fields for local $1D$ reparametrizations, local
${\cal N}=4$ supersymmetry and local $SU(2)$ R-symmetry, also some auxiliary fields. One of them (the field $D$) is a Lagrange multiplier for the important bosonic
constraint (see \p{Lagrbos}) ensuring the correct number of physical bosonic fields in the theory.}. The vanishing of these currents is associated with the local worldline
${\cal N}=4, 1D$ supersymmetry of the models considered and so this property should be equally valid for the most general ${\cal N}=4$ QK mechanics model which also
respects this local supersymmetry and the superfield action of which (as well as the bosonic component action) were given in \cite{Ivanov:2017ajf}. For the same reason,
the hamiltonian analysis of Sect. 8 should also be directly applicable  to the general case. We hope to address, from this point of view, examples of more
general QK mechanics models (with the reduced isometry groups and ${\cal L}^{+4} \neq 0$) elsewhere. An important property of ${\cal N}=4$ QK mechanics models is the possibility to add, to
the sigma-model type action, the locally ${\cal N}=4$ supersymmetric Wess-Zumino term. It would be interesting to see how the inclusion of such terms
(even in the simplest $\mathbb{H}{\rm H}^n$ or $\mathbb{H}{\rm P}^n$ cases) will affect the analysis carried out in the present paper. In   \cite{Ivanov:2017ajf}
we conjectured the simplest harmonic superspace actions describing QKT (``Quaternion-K$\ddot{a}$hler with Torsion'') [22] ${\cal N}$ = 4
mechanics models. It would be desirable to extend our analysis to this non-trivial case too.

\section*{Acknowledgements}
E.I. thanks S. Fedoruk and D. Sorokin for enlightening comments. His work was partly supported by Russian Foundation for Basic Research, project No 18-02-01046, and
by Ministry of Science and High Education of Russian Federation, project No FEWF-2020-0003. L.M. thanks
BLTP JINR for kind hospitality at the earlier stage of this study.

\providecommand{\href}[2]{#2}\begingroup\raggedright
\endgroup

\end{document}